\documentclass[aps,prd,letterpaper,showpacs,twocolumn,preprintnumbers,amsmath,amssymb,nofootinbib]{revtex4}
\usepackage{graphicx}
\newcommand{\mys}[1]{\section{#1}
    \setcounter{equation}{0}}
    \renewcommand{\theequation}{\arabic{section}.\arabic{equation}}
\newcommand{\myappendix}{\appendix
   \renewcommand{\theequation}{\Alph{section}.\arabic{equation}}
   }

\DeclareMathAlphabet   {\mathsc}{OT1}{cmr}{m}{sc}

\def\[{\left [}
\def\]{\right ]}
\def\({\left (}
\def\){\right )}
\newcommand{\lang}{\left\langle}
\newcommand{\rang}{\right\rangle}
\newcommand{\lbr}{\left\{}
\newcommand{\rbr}{\right\}}
\newcommand{\oline}[1]{\overline{#1}}

\newcommand{\wtd}[1]{\widetilde{#1}}

\newcommand{\wh}[1]{\widehat{#1}}

\newcommand{\GeV}      {~\mathrm{GeV}}
\newcommand{\TeV}      {~\mathrm{TeV}}

\newcommand{\SM}       {\mathsc{sm}}

\newcommand{\EM}       {\mathsc{em}}

\newcommand{\GUT}      {\mathsc{gut}}

\newcommand{\diff}{\mbox{d}}

\newcommand{\gappeq}{\mathrel{\rlap {\raise.5ex\hbox{$>$}}
{\lower.5ex\hbox{$\sim$}}}}
\newcommand{\lappeq}{\mathrel{\rlap{\raise.5ex\hbox{$<$}}
{\lower.5ex\hbox{$\sim$}}}}

\newcommand{\DQ}{D^\mathsc{DQ}}
\newcommand{\LQ}{D^\mathsc{LQ}}
\newcommand{\LSP}{\mathsc{LSP}}
\newcommand{\met}{\not{\hspace{-.05in}{E_T}}}
\newcommand{\lumint}{5~fb$^{-1}$}
\newcommand{\xDl}{X_{D_L}}
\newcommand{\xDr}{X_{D_R}}
\newcommand{\xdl}{X_{d_L}}
\newcommand{\xul}{X_{u_L}}
\newcommand{\xel}{X_{e_L}}
\newcommand{\xnl}{X_{\nu_L}}
\newcommand{\xdr}{X_{d_R}}
\newcommand{\xur}{X_{u_R}}
\newcommand{\xer}{X_{e_R}}
\newcommand{\yDl}{Y_{D_L}}
\newcommand{\yDr}{Y_{D_R}}
\newcommand{\ydr}{Y_{d_R}}
\newcommand{\yur}{Y_{u_R}}
\newcommand{\zDl}{Z_{D_L}}
\newcommand{\zDr}{Z_{D_R}}
\newcommand{\zdl}{Z_{d_L}}
\newcommand{\zul}{Z_{u_L}}
\newcommand{\qel}{Q_{e_L}}
\newcommand{\qul}{Q_{u_L}}
\newcommand{\rdl}{R_{d_L}}
\newcommand{\rnl}{R_{\nu_L}}
\newcommand{\ser}{S_{e_R}}
\newcommand{\sur}{S_{u_R}}
\begin{document}


\title{Theory and Phenomenology of Exotic Isosinglet Quarks and Squarks}

\author{Junhai Kang}
\affiliation{Physics Department, University of Maryland,
College Park, MD 20742}

\author{Paul Langacker}
\affiliation{School of Natural Sciences, Institute for Advanced
Study, Einstein Drive, Princeton, NJ 08540}

\author{Brent D. Nelson} 
\affiliation{Department of Physics, Northeastern University, Boston,
MA 02115}

\date{\today}

\begin{abstract}
Extensions of the MSSM often predict the existence of new fermions
and their scalar superpartners which are vectorlike with respect to the
standard model gauge group but may be chiral under additional gauge factors.
In this paper we explore the production and decay of an important example, i.e.,
a heavy isosinglet charge $-1/3$ quark and its scalar partner, using the charge assignments
of a  $\mathbf{27}$-plet of $E_6$ for illustration. We emphasize
that, depending on the symmetries of the low energy theory, such exotic particles
may decay by the mixing of the fermion with the $d,\ s$, or $b$ quarks; may decay by
leptoquark or diquark couplings (which may nevertheless preserve a form of $R$-parity);
or may be stable with respect to renormalizable couplings but decay by higher-dimension
operators on cosmological times scales. We discuss the latter two possibilities in detail for
various assumptions concerning the relative masses of the exotic fermions, scalars,
and the lightest neutralino, and emphasize the necessity of considering the collider signatures
in conjunction with the normal MSSM processes. Existing and projected constraints from colliders, indirect experiments,
proton decay, and big bang nucleosynthesis are considered.

\end{abstract}

\pacs{12.60.Jv,04.65.+e,14.80.Ly,95.35.+d}
\maketitle

\section{Introduction} \label{intro} 
As the beginning of data-taking at the CERN Large Hadron Collider
(LHC) rapidly approaches, there has been an increase in research
into both the possible new physics opportunities as well as the
challenges facing experimenter and theorist alike. The focus of
attention has thus far been squarely on new physics of a
supersymmetric nature, in particular on the minimal supersymmetric
extension of the Standard Model (MSSM). This is undoubtedly a
well-justified approach, but in our preparatory studies of
opportunities and challenges at the LHC it is wise to sometimes take
a broader view of what the TeV~scale may reveal. There has recently
been a growing
appreciation~\cite{Binetruy:2003cy,Allanach:2004my,Datta:2005vx,Allanach:2005kz,Arkani-Hamed:2005px,Plehn:2004rp}
of the possible difficulties in connecting data to underlying
theories, focusing on extracting parameters of the MSSM Lagrangian
or distinguishing the MSSM from some other paradigm of new particle
physics. In this paper we wish to consider another possibility. We
will consider here the case in which the ``usual'' new physics
signal from the states of the MSSM is combined with that from {\em
additional} supersymmetric states accessible at the TeV~scale. In
particular, we will consider the presence of additional vector-like
pairs of $SU(3)$ triplets which are singlets under the $SU(2)$ of
the Standard Model but carry non-vanishing hypercharge (as well as
possible additional $U(1)'$ charges).

Such states, which we will denote as $D$ and $D^c$ in this work, are
among the most well-motivated extensions of the MSSM particle
content. For example the supersymmetric $SU(5)$, $SO(10)$ and $E_6$
grand unified theories (GUTs), as well as non-supersymmetric $E_6$,
predict the existence of such
states~\cite{Langacker:1980js,Hewett:1988xc}, though only in the
$E_6$ case (or in nonminimal versions of   $SU(5)$ and $SO(10)$) can
some of them have electroweak-scale masses without severely
compromising the lifetime of the proton. Simple enlargements of the
electroweak sector of the MSSM incorporate these states, to saturate
anomaly cancelation constraints in $U(1)'$ models and/or to
contribute to the radiative breaking of the extended
symmetries~\cite{Cvetic:1995rj,Cvetic:1997ky,Keith:1997zb,Langacker:1998tc,Daikoku:2000ep,Erler:2000wu}.
$U(1)'$ extensions\footnote{Such extensions are consistent with
simple gauge coupling unification if the $D$ and $D^c$ occur in
combination with additional vector-like Higgs or lepton doublet
pairs, i.e., so that the new states have the quantum numbers of a
$\mathbf{5}+\mathbf{5}^*$ of $SU(5)$.} may arise, for example, as
variations on the NMSSM~\cite{Ellis:1988er} for generating dynamical
$\mu$ terms. Structures of these sorts arise routinely in top-down
string constructions, particularly those of the heterotic
string~\cite{Antoniadis:1986rn,Kawai:1986va,Kawai:1986vd,Antoniadis:1989zy,Faraggi:1993pr,Chaudhuri:1994cd,Cleaver:1998gc,Cleaver:1999cj,Giedt:2001zw}
-- so much so, in fact, that the $E_6$-version of $D$ and $D^c$ are
often referred to simply as ``superstring-inspired''
exotics.\footnote{In explicit string constructions one finds that
additional triplets/antitriplets of $SU(3)$ often have $U(1)'$
charges that distinguish these states from their GUT counterparts.}
Yet more significant (for our purposes) than these motivations is
the simple fact that the existence of additional vector-like pairs
of $SU(3)$ triplets is a very real logical possibility that will
have a significant impact on the experimental environment at a
hadron collider such as the LHC.\footnote{In this regard the present
work serves as a complement to similar treatments of new $SU(2)$
doublets and new Higgs
singlets~\cite{Ellis:1988er,nMSSM,Panagiotakopoulos:2000wp,Menon:2004wv,Dermisek:2005ar,Ellwanger:2005uu,Barger:2006dh,O'Connell:2006wi,Barger:2006sk,Dermisek:2007yt,Barger:2007im}
or other types of new physics~\cite{Barger:2007ay} at the LHC.}

Consider, for example, the various logical possibilities for
interaction between exotic $D$ and $D^c$ states and the fields of
the Standard Model. The simplest scenario, and the one that is
usually studied, especially in the non-supersymmetric case, is one
in which the exotic $D$ and the Standard Model $d$ are allowed (by
the conserved quantum numbers) to mix, as are the conjugate fields
$D^c$ and $d^c$. Another scenario might forbid this sort of mixing
(e.g., because of R-parity conservation), but allow renormalizable
interactions between the new $D$, $D^c$ and Standard Model fields in
the superpotential. Depending on the quantum numbers of the exotic
states a range of possible production and decay mechanisms would
then be operative, including interactions that are leptoquark or
diquark in nature. Still another possibility is  that some new
symmetry -- or combination of symmetries -- forbids these
renormalizable interactions but allows higher-order interactions in
the superpotential. In this case the new exotics would be
``quasi-stable''; they would have a lifetime that implies stability
on collider timescales (approximately 100ns or longer) but might
decay sufficiently quickly to avoid cosmological limits on such
objects. Finally, we mention the possibility that the exotic
particle is absolutely stable or stable on the time scale of the age
of the universe. This would imply cosmological difficulties and is
also severely constrained by searches for heavy stable particles,
e.g., in sea water (for a review, see~\cite{Perl:2001xi}). However,
these difficulties could possibly be avoided if the reheating
temperature of the universe after inflation was very
low~\cite{Giudice:2000ex}, especially if the mass is larger than a
TeV~\cite{Kudo:2001ie}.

In this work we will therefore cover a variety of such phenomena:
new signatures and new challenges for connecting data to theories,
cosmological issues of long-lived heavy $SU(3)$-charged states, and
indirect constraints coming from limits on rare processes. The goal
is to be as comprehensive as possible in enumerating all the
logically distinct ways in which these exotic quarks and squarks can
manifest themselves in Nature. In each case we wish to ask what are
the constraints on the existence of these new states, and by what
observational methods will we infer their existence? We would also
like to know how the answers to these questions will depend on the
free parameters of theories which contain such exotics. To this end
we will need to consider an array of possibilities instead of any
one paradigm. Nevertheless, to be concrete, one needs to discuss
phenomena in terms of a model or class of models. We will therefore
choose representative ones when necessary, but we strive to treat
everything as model-independently as is possible. We note that
several specific models that fall into this class have already been
studied at length, but other aspects which we cover have hardly been
mentioned in the literature. More complete references to existing
studies will be provided in the appropriate sections.

The rest of the paper is organized as follows. In
Section~\ref{cases} we will provide some background on $E_6$-based
models and explain the various ways that exotic isosinglet quarks
and squarks can arise. Much of this material will be a review of
previous work. In Section~\ref{masses} we will provide five
benchmark scenarios defined by the masses of the exotic fermion and
its scalar superpartners. These benchmark cases are designed to
cover a range of likely soft supersymmetry breaking mass scales and
subsequent phenomenology. In Section~\ref{production} we take a
first look at production processes for these exotics at hadron
colliders, and discuss the modifications we made to the {\tt PYTHIA}
event generator to handle these new states. Section~\ref{bounds}
summarizes the bounds on exotic $SU(3)$-charged isosinglets arising
from direct searches, rare flavor-changing processes and cosmology.
The latter are constraining only for cases in which the exotic
states are quasi-stable. In Section~\ref{qs} we look at the collider
phenomenology of this quasi-stable case in great detail, and give
the discovery reach for such states at the LHC. Section~\ref{decay}
is devoted to the collider phenomenology of scenarios in which the
exotic decays promptly in the detector through renormalizable
interactions. For technical reasons we will here focus on the more
tractable case in which the exotic has leptoquark-type couplings,
reserving the diquark case for a separate work. Additional material
is contained in two appendices.

\section{Outline of Cases} \label{cases} 

In this section we will present some basic model concepts that will
allow us to treat the phenomenology of exotic $D$ and $D^c$ quarks
in a semi-unified manner. The smallest extension of the MSSM capable
of incorporating all of the phenomena mentioned in
Section~\ref{intro} is motivated by $E_6$ GUT symmetries. Systems
based on the $E_6$ gauge group (or its various subgroups), with
matter states arising from the fundamental $\mathbf{27}$
representation, have been in the past -- and continue to be now --
an appealing framework for organizing models of beyond the MSSM
physics~\cite{Ellis:1985yc,Zwirner:1988mu,Volkas:1988zm,Suematsu:1996bv,Erler:2002pr,Kartavtsev:2004cf,Kang:2004bz,Kang:2004ix,King:2005jy}.
From within this framework many interesting limits and sub-models
can be studied: quasi-stable exotics, exotics which mix with SM
fermions, promptly decaying leptoquarks, promptly decaying diquarks,
and so forth. Let us emphasize that we are not interested in any
particular $E_6$ model, nor do we even insist that all elements of
grand unification are present.\footnote{For example, if the $D$ and
$D^c$ states related to the Higgs doublets responsible for fermion
mass generation are at the TeV scale (as is required if there is a
TeV scale $U(1)'$ gauge symmetry), then the $E_6$ relations between
the Higgs and $D,D^c$ Yukawa couplings must not be respected in
order to avoid rapid proton decay. Fortunately, string constructions
often do not honor such relations.} We will also not concern
ourselves much with the additional richness that such models present
(new $Z'$-bosons, new Higgs doublets, additional neutralinos and Higgs singlets, challenges for neutrino mass
generation and gauge coupling unification, etc.). We merely
introduce some basic facts of $E_6$-inspired models for the sake of
coherence of presentation and as an example of consistent,
anomaly-free exotic particle quantum numbers that can be consistent
with gauge unification. Many constructions which do not fit into the
$E_6$ framework can give rise to the physics we will describe in
Sections~\ref{qs} and~\ref{decay}, and our discussion of the
consequences of exotic $D$ and $D^c$ pairs will be quite general.

\subsection{General Framework} \label{genE6}

The $E_6$ gauge group decomposes to the gauge group of the Standard
Model as
\begin{eqnarray}
E_6 &\to& SO(10) \times U(1)_{\psi} \nonumber \\
 &\to& SU(5) \times U(1)_{\chi} \times U(1)_{\psi} \label{E6break} \\
 &\to& SU(3) \times SU(2) \times U(1)_Y
 \times U(1)_{\chi} \times U(1)_{\psi}, \nonumber
\end{eqnarray}
where the designation of the particular $U(1)$ combinations
$U(1)_{\psi}$ and $U(1)_{\chi}$ are traditional and chosen for
convenience of notation. The above symmetry breaking can occur by a
variety of physical means. For our purposes the exact mechanism is
unimportant, though we might imagine a string-theoretic origin for
our exotics, in which case a Wilson line breaking analogous to the
Hosotani mechanism may be envisioned~\cite{Witten:1985xc}. There are
two additional $U(1)$ factors beyond those of the Standard Model.
One or two linear combinations of these factors may remain intact to
very low energies, though both must be broken at the electroweak
scale. We will assume the associated $Z'$-bosons are heavy
enough to be
irrelevant to the LHC phenomenology we wish to explore.

\begin{table}
{\begin{center}
\begin{tabular}{c||cccc}
\parbox{1.4cm}{Field} & \parbox{1.4cm}{$Q_Y$} &
\parbox{1.4cm}{$2\sqrt{6}Q_{\psi}$} &
\parbox{1.4cm}{$2\sqrt{10}Q_{\chi}$}&
\parbox{1.4cm}{$2\sqrt{10}Q_{N}$} \\ \hline \hline
$Q_i$ & 1/6 & 1 & -1& 1 \\ \hline
${u}_i^c$ & -2/3 & 1 & -1& 1 \\ \hline
${d}_i^c$ & 1/3 & 1 & 3& 2 \\ \hline
$L_i$ & -1/2 & 1 & 3& 2 \\ \hline
${e}_i^c$ & 1 & 1 & -1 & 1\\ \hline
${\nu}_i^c$ & 0 & 1 & -5 & 0\\ \hline
$(H_u)_i$ & 1/2 & -2 & 2& -2 \\ \hline
$(H_d)_i$ & -1/2 & -2 & -2& -3 \\ \hline
$D_i$ & -1/3 & -2 & 2& -2 \\ \hline
$D^c_i$ & 1/3 & -2 & -2& -3 \\ \hline
$S_i$& 0 & 4 & 0& 5 \\ \hline
\end{tabular}
\end{center}}
{\caption{\label{tbl:27} {\bf Decomposition of the fundamental of
$E_6$ under the Standard Model gauge group}. The fields of the
$\mathbf{27}_i$ representation of $E_6$ are here given in terms of
their $SU(3)\times SU(2)$ representation, as well as their charges
under the four $U(1)$ factors. ($Q_N$ is relevant to the
quasi-stable scenario.) The index $i=1,\;2,\;3$ represents a
generation index. Some of the $H_{u,d}$ pairs could be interpreted as
exotic lepton doublets if they do not acquire expectation values.}}
\end{table}

Each fundamental representation $\mathbf{27}$ of $E_6$ contains the
Standard Model representations given in Table~\ref{tbl:27}. The
particle content of the $\mathbf{16}$ representation of $SO(10)$ is
augmented by a pair of Higgs doublets, a pair of exotic quarks $D$
and $D^c$ and a singlet field $S$. This particle content is
anomaly-free by construction. Achieving three generations of
Standard Model fields therefore implies three generations of Higgs
fields, exotic triplets/antitriplets and singlets. With the field
content of Table~\ref{tbl:27} it is well known that gauge coupling
unification cannot be achieved consistent with the measured
low-energy values of $\alpha_3$, $\alpha_{\EM}$ and
$\sin^2\theta_W$. This can be easily remedied by the introduction of
additional fields in an anomaly-free
manner~\cite{Gaillard:1992yb,Martin:1995wb,Langacker:1998tc}. As
this is immaterial to our purposes we will not consider the issue
further.

The purpose of introducing the field content of the $\mathbf{27}$
representation was simply to motivate the form that superpotential
interactions of $D$ and $D^c$ might be allowed to take. The
renormalizable superpotential for $E_6$ is simply given by $W =
\lambda_{ijk} \mathbf{27}_i \mathbf{27}_j \mathbf{27}_k$. However,
we do not insist  on full $E_6$ invariance of the superpotential,
but rather use this form to motivate the classes of allowed
couplings. When decomposed into the appropriate components
under~(\ref{E6break}) $W$ then becomes
\begin{eqnarray}
W &=& \lambda^1_{ij} Q_i u^c_j H_u + \lambda^2_{ij} Q_i d^c_j H_d +
\lambda^3_{ij} L_i e^c_j H_d \nonumber \\ & & + \lambda^{11}_{ij}
L_i \nu^c_j H_u + \lambda^4 S H_u H_d + \lambda^5_{ij} S D_i
D^c_j \nonumber \\ & & + W_{\rm LQ} + W_{\rm DQ}
\label{W} \end{eqnarray}
where
\begin{eqnarray}
W_{\rm LQ} = \lambda^6_{ijk} D_i u^c_j e^c_k + \lambda^7_{ijk} D^c_i
Q_j L_k + \lambda^8_{ijk} D_i d_j^c \nu^c_k \label{WLQ} \\
W_{\rm DQ} = \lambda^9_{ijk} D_i Q_j Q_k + \lambda^{10}_{ijk} D^c_i
u^c_j d^c_k , \label{WDQ}
\end{eqnarray}
and the convention for numbering the interactions is taken from the
review of Hewett and Rizzo~\cite{Hewett:1988xc}. We have restricted
our attention here to one relevant generation of each. This could be
achieved by appropriate assumptions concerning the Yukawa matrices
and/or the vacuum expectation values (vevs) of the singlet fields
$S_i$. We will usually restrict our attention to a single
generation of $D$ and $D^c$ quarks as well, but for now we retain
the generation label.

If we were to demand invariance {\em only} under the Standard Model
gauge group then additional
 bilinear and trilinear terms (e.g., terms involving just the standard model fields
 which violate R-parity) could be
added to~(\ref{W})~\cite{Rizzo:1992ts}.\footnote{Fundamental
bilinears are not allowed by $E_6$ gauge invariance with only
fundamental $\mathbf{27}$ representations. Furthermore, if we
imagine a string-theoretic origin for our exotic $D$ and $D^c$
states then such terms are generally forbidden if these fields are
to be considered part of the massless spectrum of the string.} If we
require invariance under only the Standard Model plus {\em one}
additional $U(1)$ factor then some subset of these additional terms
may be allowed. If one additional $U(1)$ factor arising from the
original $E_6$ is retained to low energies it is traditionally
parameterized as
\begin{equation}
Q' = Q_{\chi}\cos\theta_E + Q_{\psi}\sin\theta_E , \label{Qprime}
\end{equation}
where the charges $Q_{\chi}$ and $Q_{\psi}$ are those given in
Table~\ref{tbl:27}. Any choice of $\theta_E$ in~(\ref{Qprime})
allows all the terms in~(\ref{W})-(\ref{WDQ}), by construction.
Higher-dimensional, non-renormalizable operators are also possible
in the superpotential. Their presence or absence depends on which
linear combination in~(\ref{Qprime}), if any, is assumed to be
present at low-energies.
For the sake of concreteness, when
necessary we will choose $U(1)'$ charge assignments for these fields
according to the $U(1)_{\eta}$ combination with $\theta_E = 2\pi -
\tan^{-1}\sqrt{5/3}$, or to the  $U(1)_N$ combination with $\theta_E =
\tan^{-1}\sqrt{15}$.

\subsection{Charge Assignments} \label{E6case}

If both~(\ref{WLQ}) and~(\ref{WDQ}) are present simultaneously then
it is impossible to assign an unambiguous $B$ and $L$ quantum number
to $D$ and $D^c$ -- thus $B$ and $L$ are broken. In this case the
exotic $SU(3)$-charged states will mediate rapid proton decay. We
will therefore insist on separately conserved quantum numbers $B$ and
$L$ and choose superpotential terms to allow definite $B(D)$ and
$L(D)$ assignments. This will always imply a trivially conserved
R-parity quantum number using the standard definition $R_p =
(-1)^{3(B-L) + 2s}$.

When only~(\ref{WLQ}) is present then one can assign the quantum
numbers $B(D) = 1/3$ and $L(D)=1$ so that $R_p(D) = -1$ and
$R_p(\wtd{D}) = +1$ and we can identify $D$ as a leptoquark. With
only~(\ref{WDQ}) we have $B(D) = -2/3$ and $L(D)=0$ and the same
$R_p$ assignment; the state $D$ is then a diquark. Note that in
these two cases the $D$ and $D^c$ are like $H_u$ and $H_d$: the
scalar is the ``standard'' particle and the fermion is the
``partner''. So the R-parity distinguishes $\mathbf{\bar{5}}$'s
associated with the $\mathbf{16}$ of $SO(10)$ from those coming from
the $\mathbf{10}$'s of $SO(10)$.  In this case the only renormalizable operators
allowed are those of~(\ref{W}) with~(\ref{WLQ}) {\em
or}~(\ref{WDQ}). All dimension-five operators involving the exotic
$D$ and $D^c$ also vanish in this case.

If we instead insist on baryon and lepton number conservation with
$B(D) = 1/3$ and $L(D)=0$, then the exotic $D^c$ has the same baryon
and lepton number as the Standard Model $d^c$ field. Now $R_p(D) =
+1$ and $R_p(\wtd{D}) = -1$ as with the quarks of the Standard
Model. In this case {\em both}~(\ref{WLQ}) {\em and}~(\ref{WDQ}) are
forbidden, leaving only the first two lines of~(\ref{W}). At the
renormalizable level, therefore, this is an accidentally conserved
quantum ``$D$-number'' for the exotic fields and they are stable. Operators
connecting the fields $D$ and $D^c$ to the Standard Model may be
allowed at the non-renormalizable level, however, depending on the
$U(1)'$ charge assignments.\footnote{The possibility of
strongly interacting or charged exotics that
are absolutely stable, or which are stable on the time scale of the age of the universe, was
commented on in the Introduction.}
 In particular, for the case of the $U(1)_N$ combination, where $\theta_E =
\tan^{-1}\sqrt{15}$~\cite{Ma:1995xk,Kang:2004ix,King:2005jy}, the
combination of $B$, $L$, and $U(1)_N$ symmetry forbids the
renormalizable operators beyond the first two lines of~(\ref{W}),
but allows the dimension-five operators
\begin{eqnarray}
{\rm dim5}&:& D^c Q H_d S, \; D^c Q Q u^c, \; D^c Q L \nu^c ,
\label{Wdim5}
\end{eqnarray}
which preserve R-parity. These, along with the term proportional to
$\lambda^5$ in~(\ref{W}) (which leads to a $D, D^c$ mass), allow for
the decay of the exotics $D$ and $D^c$, which are therefore
quasi-stable. An alternative model of quasi-stable exotics, in which
a $U(1)'$ gauge symmetry alone forbids $D$ decay at the
renormalizable level, can be found in Appendix~\ref{qsalt}.

Finally, the case of mixing between the exotic $D$ and SM $d$-quark
leads to decays such as $D\rightarrow uW$, $D\rightarrow dZ$, and in
some cases to $D\rightarrow dZ'$, or $D \rightarrow d+$ Higgs, where
the $W,\ Z,\ Z'$, or Higgs can be real or virtual. Such mixing can
be induced by the operator $\lambda^8D d^c  \nu^c$  in the presence
of a sneutrino vev. Such examples are often considered in the case
of extensions of the MSSM in which one assigns $L(\nu^c)=0$, as is
often put forward in rank-6 models. Mixing can also be induced by
the operator $\lambda^7D^c Q L$ if the scalar component of the
neutrino in $L$ acquires a vev, or by other operators such as $ D^c
Q H_d$ that are not included in~(\ref{W}) because they don't occur
in the singlet of $\mathbf{27}^3$ and therefore violate the extra
$U(1)$ symmetries. The case of mixing between exotics and SM quarks
(through arbitrary mechanisms) and its phenomenology has been
well-covered in the
literature~\cite{Barger:1985nq,Langacker:1988ur,Andre:2003wc,Mehdiyev:2006tz},
especially in the non-supersymmetric case. We
will therefore focus on cases where such a sneutrino vev or other
mechanism is absent for the rest of this work.

\subsection{Mass Patterns} \label{masses}

We list in~(\ref{W}) the operators $\lambda^4$ and $\lambda^5$ for
completeness, but they are not particularly relevant for our study.
The field $S$ is a singlet under the Standard Model gauge group, but
generally carries charges under additional $U(1)$'s which arise from
the breaking of $E_6$ to the Standard Model. A vev for this field
would generate both an effective $\mu$ parameter as well as a
supersymmetric mass for the $D$ and $D^c$. It will also generally
break one or more additional $U(1)$'s, producing new (heavy)
$Z'$-bosons. As these facts are not relevant to the phenomenology we
will pursue in subsequent sections, we will not consider them
further. We do note, however, that the scalar mass matrices for the
$\wtd{D}$ and $\wtd{D}^c$ will generally depend on the charges of
these fields under any additional $U(1)'$ through various D-terms.

Like the squarks and sleptons of the MSSM, the exotic scalar sector
can also be described by a $6\times 6$ scalar mass matrix. Unlike
the MSSM fields, however, the fermionic modes are not protected by
Standard Model chiral symmetries from receiving large masses. In
fact, we generally expect the scalar and fermionic components to
receive a common, supersymmetric mass determined by the vev of some
field such as the singlets $S_i$ in~(\ref{W}). Let us for the moment
restrict our attention to the case of one generation of exotics.
Defining $ \lang S \rang\equiv s$ and keeping in mind the
definitions $\lambda^4 s = \mu_{\rm eff}$ and $\lambda^5  s = m_D$
we can write the scalar mass matrix as
\begin{equation}
m_{\wtd{D}}^2 = \( \begin{array}{cc} m_{aa}^2 & m_{ab}^2 \\
m_{ab}^2 & m_{bb}^2 \end{array} \)
\label{mDsq} \end{equation}
with
\begin{eqnarray}
m_{aa}^2 & = & m_{\wtd{D}}^2 + m_D^2 + \frac{1}{3}\sin^2
\theta_W\cos 2\beta M_Z^2 \nonumber \\ & &+g'\,^2 Q'_D
(Q'_S s^2 + Q'_{H_u} v_u^2 + Q'_{H_d} v_d^2) \nonumber \\
m_{bb}^2 & = & m_{\wtd{D}^c}^2 + m_D^2 - \frac{1}{3}\sin^2
\theta_W\cos 2\beta M_Z^2 \nonumber \\ & &+g'\,^2
Q'_{D^c} (Q'_Ss^2 + Q'_{H_u} v_u^2 + Q'_{H_d} v_d^2) \nonumber \\
m_{ab}^2 &=& m_D \(A_5 + \mu_{\rm eff}\(\frac{v_u v_d}{s^2}\)\) ,
\label{mDentries}\end{eqnarray}
where $\lang H^0_{u,d} \rang \equiv v_{u,d}$ are the usual scalar
Higgs vevs and $\tan \beta = v_u/v_d$. The quantity $A_5$ is the
soft supersymmetry-breaking A-term associated with the Yukawa
interaction $\lambda^5$ in~(\ref{W}). We have ignored possible
CP-violating phases. In this work, equations~(\ref{mDsq})
and~(\ref{mDentries}) are the only places where the precise $Q'$
charges of the fields will be required. Note that the D-term
contributions to $m_{aa}^2$ and $m_{bb}^2$ are typically small
perturbations when $g' \simeq g_Y$, where $g_Y \equiv \sqrt{5/3}
g_1$ is the GUT-normalized $U(1)_Y$ coupling, and can be absorbed
into the values of the soft masses $m_{\wtd{D}}^2$ and
$m_{\wtd{D}^c}^2$ at low energies. The resulting masses for the
$U(1)_{\eta}$ model and $U(1)_N$ model are given for five benchmark
points in Table~\ref{benchmarks}. We will use these benchmark points
to illustrate aspects of collider phenomenology in Sections~\ref{qs}
and~\ref{decay} below.

\begin{table}
{\begin{center}
\begin{tabular}{|c||c|c|c|c|c|} \hline
\parbox{1.6cm}{Parameter} & \parbox{1.2cm}{A} &
\parbox{1.2cm}{B} & \parbox{1.2cm}{C} &
\parbox{1.2cm}{D} & \parbox{1.2cm}{E} \\
\hline \hline
$M_{D_{1/2}}$ & 300 & 300 & 300 & 600 & 1000 \\
$m_{D_{0}}$ & 400 & 400 & 1000 & 400 & 400 \\
$m_{D_{0}^c}$ & 400 & 400 & 1000 & 400 & 400 \\
$A_5$ & 350 & 150 & 100 & 600 & 1050 \\ \hline
\multicolumn{1}{c}{} & \multicolumn{5}{c}{$U(1)_{\eta}$ Model} \\
\hline
$M_{D_0^1}$ & 367 & 441 & 1024 & 388 & 318 \\
$M_{D_0^2}$ & 587 & 553 & 1053 & 932 & 1482 \\ \hline
\multicolumn{1}{c}{} & \multicolumn{5}{c}{$U(1)_{N}$ Model} \\
\hline
$M_{D_0^1}$ & 360 & 435 & 1022 & 381 & 309 \\
$M_{D_0^2}$ & 582 & 527 & 1050 & 929 & 1480 \\ \hline
\end{tabular}
\end{center}}
{\caption{\label{benchmarks} {\bf Sample spectra for the exotic SUSY
sector}. Five benchmark mass patterns designed to illustrate the
possible collider signatures of exotic supermultiplets. All values
are in GeV at the electroweak scale. These examples will be used
extensively in Sections~\ref{qs} and~\ref{decay} below. }}
\end{table}

For a particular set of $U(1)'$ charges we can define the mass
splittings
\begin{eqnarray}
\Delta_1 &\equiv& m_{D_{1/2}} - m_{D_0^1} \label{delta1} \\
\Delta_2 &\equiv& m_{D_{1/2}} - m_{D_0^2} \label{delta2}
\end{eqnarray}
between the physical fermion $D_{1/2}$ (with mass $m_{D_{1/2}} =
m_D$) and the lightest scalar $D_0^1$ or heavier scalar $D_0^2$,
respectively. These splittings are functions of the dimensionful
parameters $m_D$, $m_{\wtd{D}}$, $m_{\wtd{D}^c}$, $A_5$ and
$\mu_{\rm eff}$, as well as the dimensionless parameters $g'$,
$\tan\beta$ and $\lambda^4$ (or alternatively the ratio $ v_u v_d
/s^2$). If the quantity $\Delta_1$ is positive, then the lightest
exotic particle (LEP) is the scalar. If it is negative then the LEP
is the fermion.\footnote{Note that with the conventions
of~(\ref{delta2}) the quantity $\Delta_2$ will generally be
negative.} As an example of the types of mass hierarchies that are
possible in this parameter space, let us take the charges of the
$U(1)_{\eta}$ model with $g'=g_Y$ at the electroweak scale. Let us
also fix the value of $\tan\beta =10$, $\mu_{\rm eff} = 370 \GeV$
and $\lambda^4 = 0.5$ (implying $s= 740 \GeV$).

\begin{figure}[tb]
\begin{center}
\includegraphics[scale=0.6,angle=0]{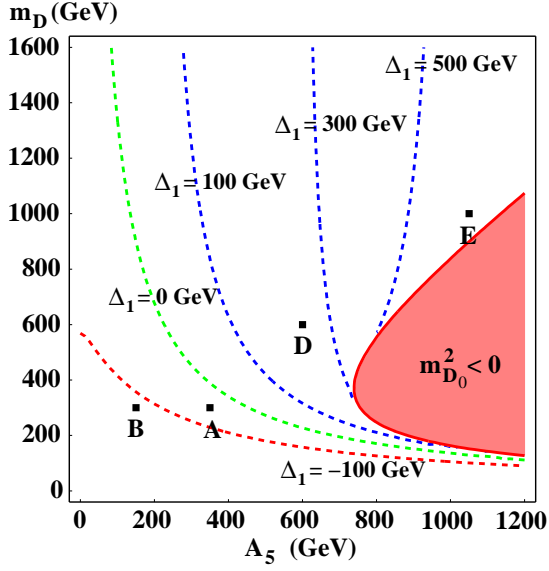}
\caption{\footnotesize \textbf{Mass hierarchy $\Delta_1$ for fixed
soft scalar masses.} The quantity $\Delta_1 \equiv m_{D_{1/2}} -
m_{D_0^1}$ is plotted as a function of the fermion mass $m_D$ and
the soft supersymmetry-breaking trilinear $A_5$ for fixed (common)
soft scalar masses $m_{D_0} = m_{\wtd{D}} = m_{\wtd{D}^c} = 400
\GeV$. The shaded region in the lower right produces a negative
mass-squared for the lightest scalar exotic. Relevant points from
Table~\ref{benchmarks} are labeled.} \label{fig:400diff1}
\end{center}
\end{figure}

\begin{figure}[t]
\begin{center}
\includegraphics[scale=0.6,angle=0]{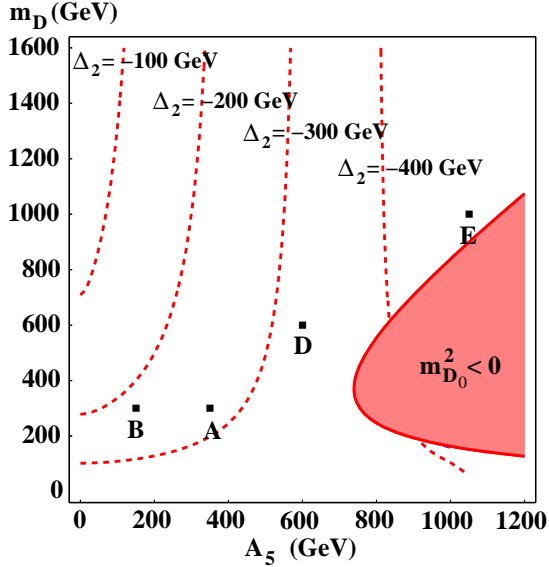}
\caption{\footnotesize \textbf{Mass hierarchy $\Delta_2$ for fixed
soft scalar masses.} Same as Figure \ref{fig:400diff1} for the
quantity $\Delta_2 \equiv m_{D_{1/2}} - m_{D_0^2}$.}
\label{fig:400diff2}
\end{center}
\end{figure}

Contours of $\Delta_1$ and $\Delta_2$ for fixed values of the
(common) scalar mass $m_{D_0} = m_{\wtd{D}} = m_{\wtd{D}^c} = 400
\GeV$ are shown in Figures~\ref{fig:400diff1}
and~\ref{fig:400diff2}, respectively. The shaded regions in the
lower right of the plots are theoretically excluded in that they
produce a tachyonic eigenvalue of the exotic scalar mass matrix. For
these low values of the soft scalar masses the majority of the
parameter space admits a hierarchy in which the scalar is the LEP.
In the limit as $m_D \to 0$ (as is effectively the case for the
fermions of the Standard Model), or in the limit where $A_5 \to 0$,
the hierarchy is such that the scalars are generally heavier than
the fermions. This region has a lower bound dictated by direct
searches limits on the mass of exotic $SU(3)$ charged fermions,
which we will discuss in the next section. For convenience we have
labeled the relevant benchmark points from Table~\ref{benchmarks}.

\begin{figure}[t]
\begin{center}
\includegraphics[scale=0.6,angle=0]{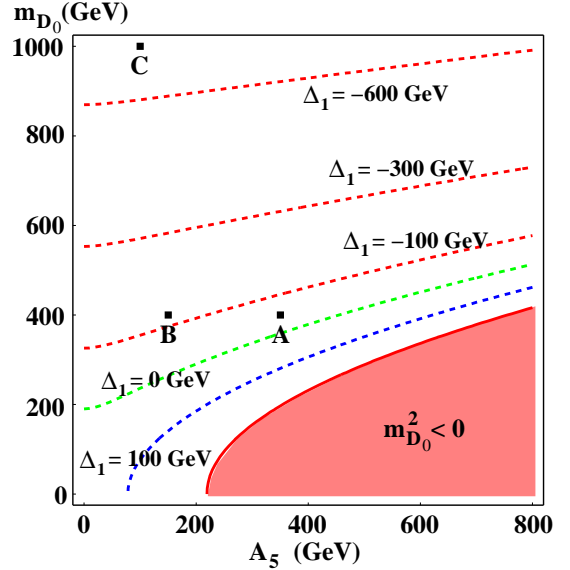}
\caption{\footnotesize \textbf{Mass hierarchy $\Delta_1$ for fixed
supersymmetric fermion masses.} The quantity $\Delta_1 \equiv
m_{D_{1/2}} - m_{D_0^1}$ is plotted as a function of the (common)
soft scalar mass $m_{D_0} = m_{\wtd{D}} = m_{\wtd{D}^c}$ and the
soft supersymmetry-breaking trilinear $A_5$ for fixed fermion mass
$m_D = 300 \GeV$. The shaded region in the lower right produces a
negative mass-squared for the lightest scalar exotic.}
\label{fig:300diff1}
\end{center}
\end{figure}

\begin{figure}[h]
\begin{center}
\includegraphics[scale=0.6,angle=0]{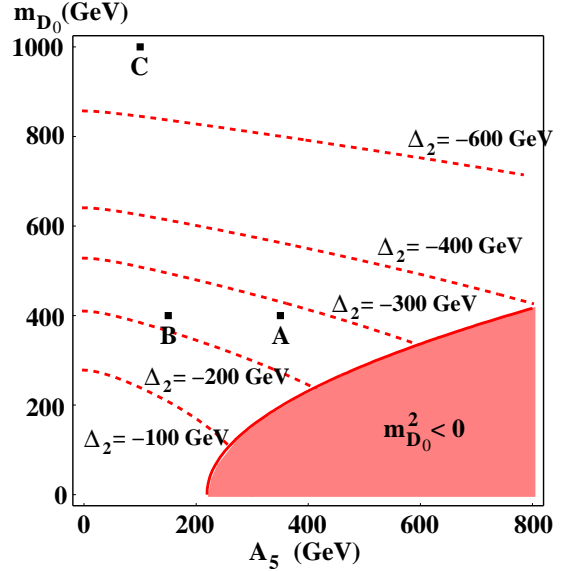}
\caption{\footnotesize \textbf{Mass hierarchy $\Delta_2$ for fixed
supersymmetric fermion masses.} Same as Figure \ref{fig:300diff1}
for the quantity $\Delta_2 \equiv m_{D_{1/2}} - m_{D_0^2}$.}
\label{fig:300diff2}
\end{center}
\end{figure}

In Figures~\ref{fig:300diff1} and~\ref{fig:300diff2} the same pair
of quantities are plotted but with the supersymmetric fermion mass
$m_D$ held fixed at 300~GeV. Again, the region in the lower right is
excluded from theoretical grounds. From these plots it is clear that
the fermion is the LEP unless the soft scalar masses are smaller
than (or on the order of) the fermion mass and/or the trilinear
$A$-term coupling is sufficiently large.

\subsection{Production at Hadron Colliders} \label{production}

We expect strongly-interacting particles with the masses given in
Table~\ref{benchmarks} to be produced relatively frequently at
hadron colliders. In this section we will discuss the production
cross-section for the leptoquark and diquark cases before
considering the direct search limits in the next section. Some
aspects of the production of exotic $SU(3)$-charged states have been
considered
elsewhere~\cite{Andre:2003wc,Mehdiyev:2006tz,Hewett:1987yg,Blumlein:1996qp,Dion:1997jw,Eboli:1997fb,Dion:1998wr},
at varying levels of sophistication and approximation.

As our goal is to be as complete as possible, we will consider the
following ten $2\to 2$ production processes: $q \, \bar{q} \to
D_{1/2} \oline{D}_{1/2}$, $g \, g \to D_{1/2} \oline{D}_{1/2}$, $q
\,\bar{q} \to D_{0} \oline{D}_{0}$, $g \, g \to D_{0} \oline{D}_{0}$
and $q \, g \to D_{0} + f$ (and c.c.), with five each for the
leptoquark and diquark cases. In addition, the couplings of $W_{\rm
DQ}$ in~(\ref{WDQ}) allow for the possibility of resonant production
of exotic diquark scalars through quark or anti-quark annihilation.
Where unavailable in the literature (or where available expressions
were incomplete) we have computed the relevant parton-level
production cross-sections to leading order and checked them against
the results of {\tt CompHEP}~\cite{Pukhov:1999gg}. These expressions
have been collected in Appendix~B. The numerical evaluation of these
cross-sections -- as well as all collider analysis performed in this
work -- was carried out with the {\tt PYTHIA 6.327} computer
package~\cite{Sjostrand:2003wg}. While the publicly-available
version of {\tt PYTHIA} does contain a scalar leptoquark, it does
not have its superpartner, nor the diquark cases we wish to study.
In addition, the scalar leptoquark contained in {\tt PYTHIA} does
not interact with the fields of the MSSM and can only decay into a
quark and a charged lepton. Therefore some substantial modification
to the off-the-shelf {\tt PYTHIA} package was required. We wish to
briefly describe these modifications here in this section before
proceeding. Further details of the analysis tools will be given in
Section~\ref{decay}.

Adding the desired new particles and interactions required the
modification of three existing subroutines and the addition of three
new routines. Six new particles (two scalars and a fermion for the
leptoquark and the diquark) were added to empty positions in the
relevant common blocks, specifically the {\tt PYDAT2}, {\tt PYDAT3}
and {\tt PYDAT4} common blocks. Masses and mixings of the new states
were computed using the formulae of~(\ref{mDentries}) via a new
routine which parallels that of {\tt PYTHRG} for standard MSSM
scalars. A call to this new routine was inserted into the
pre-existing {\tt PYMSIN} SUSY initialization subroutine. Decay
rates for the exotics into Standard Model and MSSM states are
computed and the necessary decay tables populated with a new
subroutine which is called from {\tt PYINIT}. We will discuss the
specific decay products considered in Section~\ref{decay} below.

\begin{figure}[t]
\begin{center}
\includegraphics[scale=0.6,angle=0]{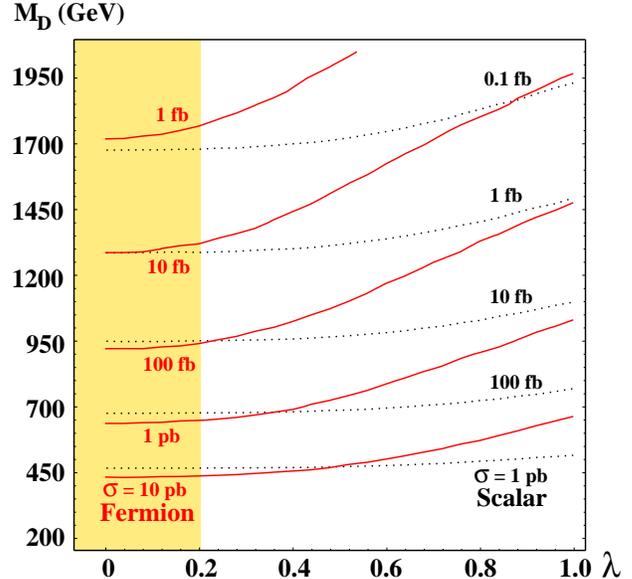}
\caption{\footnotesize \textbf{Production cross section for pairs of
leptoquarks at the LHC.} Pair production of exotic fermions ($g\,g,
q\,\bar{q} \to D\,D^c$) is given by the solid (red) contours, while
that of scalars is given by the dotted (black) contours. The region
of coupling $\lambda \lappeq 0.2$ suggested by the indirect
constraints considered in Section~\ref{bounds} is indicated by the
light shading.} \label{fig:xsecLQ}
\end{center}
\end{figure}

\begin{figure}[h]
\begin{center}
\includegraphics[scale=0.6,angle=0]{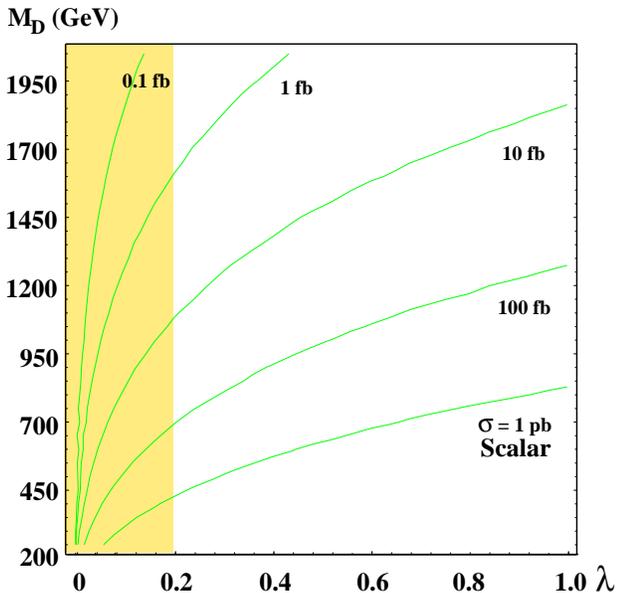}
\caption{\footnotesize \textbf{Production cross section for scalar
leptoquarks in association with fermions at the LHC.} Contours give
the production cross section for the process $q\,g \to D_0\, f$. The
region of coupling $\lambda \lappeq 0.2$ suggested by the indirect
constraints considered in Section~\ref{bounds} is indicated by the
light shading.} \label{fig:x329}
\end{center}
\end{figure}

The eleven new production processes were inserted into empty
positions in the relevant common blocks, namely {\tt PYINT2}, {\tt
PYINT4} and {\tt PYINT6}. The parton-level cross-sections were
computed in a new subroutine called from the {\tt PYSIGH} master
routine. The most significant modification of a pre-existing routine
involved {\tt PYSCAT}, which sets up the hard scattering process and
documents the color flow through the interaction. For the leptoquark
interaction, standard {\tt PYTHIA} color flow algorithms suffice,
but not so for the diquark interactions of~(\ref{WDQ}). These
vertices involve three triplets or three anti-triplets of $SU(3)$ --
an interaction not present in the Standard Model. Such cases were
not part of the original menu of color flow options in {\tt PYTHIA},
so new ones were designed and inserted into the {\tt ICOL} array for
both diquark pair production and resonant production of scalar
diquarks. The essence of these modifications was to generate
place-holding ``junctions'' to serve as sinks or sources of
color/anti-color. This modification is in the spirit of those used
to study R-parity or baryon-number violating processes in the
MSSM~\cite{Sjostrand:2002ip}.

\begin{figure}[t]
\begin{center}
\includegraphics[scale=0.6,angle=0]{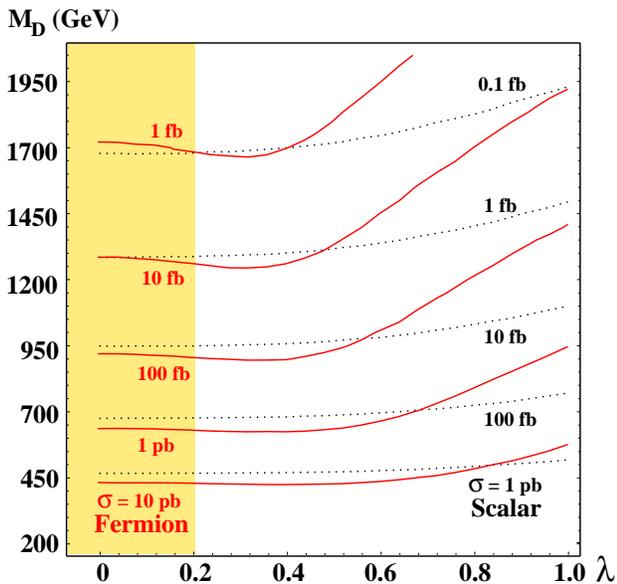}
\caption{\footnotesize \textbf{Production cross section for pairs of
diquarks at the LHC.} Same as Figure~\ref{fig:xsecLQ} but for
diquarks.} \label{fig:xsecDQ}
\end{center}
\end{figure}

\begin{figure}[h]
\begin{center}
\includegraphics[scale=0.6,angle=0]{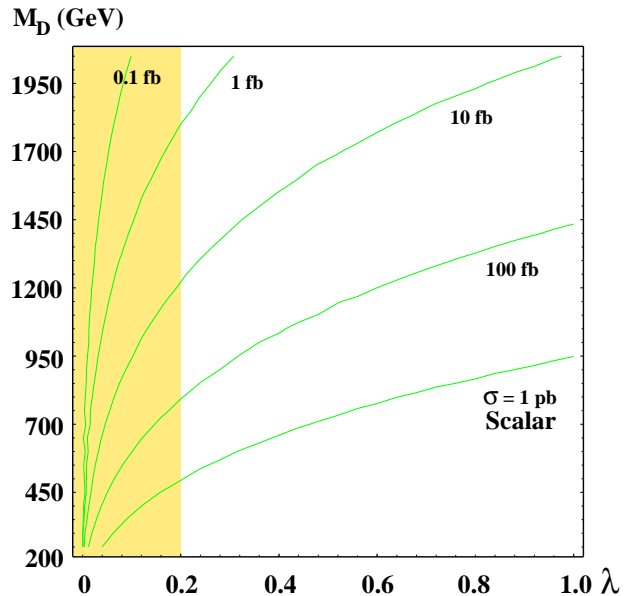}
\caption{\footnotesize \textbf{Production cross section for scalar
diquarks in association with fermions at the LHC.} Same as
Figure~\ref{fig:x329} but for diquarks.} \label{fig:x334}
\end{center}
\end{figure}

The above modifications allow us to simulate the eleven
hard-scattering processes at LHC energies. For the sake of
simplicity we will always take $\lambda^6 = \lambda^7$ and
$\lambda^9 = \lambda^{10}$ in performing simulation-based
calculations. We will refer to this common coupling as $\lambda$,
understanding that a different $\lambda$ value is implied for the
leptoquark and the diquark. Resonant production of scalar diquarks
was studied in detail elsewhere~\cite{Atag:1998xq,Cakir:2005iw}; we
postpone discussion of this case to Section~\ref{decay}. The
production cross-sections for leptoquarks are given in
Figures~\ref{fig:xsecLQ} and~\ref{fig:x329}, while those for the
diquark case are given in Figures~\ref{fig:xsecDQ}
and~\ref{fig:x334}. Pair production cross-sections of exotic quarks
and squarks are given in Figures~\ref{fig:xsecLQ}
and~\ref{fig:xsecDQ} as a function of the mass of the exotic
particle (denoted collectively as $M_D$) and the Yukawa coupling
$\lambda$. Exotic scalar production in association with a Standard
Model fermion via the process $q\,g \to D_0 \, f$ is shown in
Figures~\ref{fig:x329} and~\ref{fig:x334}. In all figures we have
shaded the region of small Yukawa coupling $\lambda \leq 0.2$. In
Section~\ref{bounds} we will see that this may be taken as a very
crude estimate of the allowed values of a typical Yukawa coupling in
this class of models. As these bounds are sensitive to many
model-dependent phenomena we have chosen to display the
cross-sections over a wide range of Yukawa parameters.

Pair production of exotic fermions via the process $q\,\bar{q} \to
D_{1/2}\, \oline{D}_{1/2}$ can proceed through $t$-channel exchange of
scalar quarks and/or scalar leptons. It is therefore necessary to
specify the masses of the superpartners of the Standard Model fields
in order to unambiguously compute the production rate at the LHC.
For the analysis presented here we will choose the well-studied
benchmark model SPS~1A from the ``Snowmass Points and Slopes''
collection~\cite{Allanach:2002nj}, in which the relevant masses are
$m_{\tilde{d}_{1}} \simeq m_{\tilde{u}_{1}} = 535\GeV$ and
$m_{\tilde{e}_{1}} = 146 \GeV$. The full set of superpartner masses
for this benchmark point will be discussed in Section~\ref{decay}
below.

The rate for production of exotic fermions is roughly an order of
magnitude larger than that for identical-mass scalars, as one
typically expects~\cite{Datta:2005vx}. The five cases in
Table~\ref{benchmarks} were specifically chosen to give at least one
exotic state in the 300-400 GeV range, ensuring a reasonable
production rate at the LHC. In fact, the total production rate of
exotics in Cases~A-C in Table~\ref{benchmarks} is roughly equivalent
to the total production rate of ``standard'' MSSM superpartners for
the SPS~1A benchmark model. This implies that it should be possible
to place meaningful limits on exotic masses and couplings from
direct searches at existing colliders. We therefore turn our
attention to direct and indirect experimental constraints on these
parameters.

\section{Current Experimental Bounds} \label{bounds} 

\subsection{Direct Search Constraints} \label{collider}

The exotic quarks $D$ and $D^c$ are charged under $SU(3)$ and (as we
demonstrated in the previous section) can thus be produced in large
numbers through QCD processes. Limits can be placed on their masses
by measuring the cross-section $\times$ branching ratio for the
exotic scalars and fermions into certain final-state topologies.
These branching fractions -- and to some extent the production rates
as well -- depend on the values of the allowed Yukawa interactions
such as those in~(\ref{W}). Here we wish to briefly summarize the
limits placed on certain manifestations of exotic $SU(3)$ triplets
from various collider searches.

For diquarks current limits extend only to scalars which decay
exclusively to pairs of jets. These jets can be produced resonantly
with an exotic scalar in the s-channel. The CDF~experiment was able
to exclude diquarks of the $E_6$ type at the 95\% confidence level
in the mass range of roughly 300 to 450 GeV by measuring the
cross-section to produce a resonant pair of dijets in a certain
invariant mass window~\cite{Harris:1995iq,Abe:1995jz,Abe:1997hm}.
For reduced branching ratios (which is the case for much of our
parameter space) the limit essentially disappears.

Leptoquarks have been more thoroughly studied at a number of
collider environments. The D0~experiment reported a limit in Run~I
of the Tevatron on the pair production of scalar leptoquarks which
decay to the final states $\nu\nu q q$, $\ell \nu q q$ and $\ell
\ell q q$ for first and second generation charged leptons. This
corresponds to a mass limit at the 95\% confidence level of $98
\GeV$ for exclusive decay to the quark/neutrino final state. For
${\rm Br}(D_0 \to \ell q) =1$ the reported limit was $m_{D_0^1}
\gappeq 200 \GeV$ and for ${\rm Br}(D_0 \to \ell q) =0.2$ a limit of
$m_{D_0^1} \gappeq 150 \GeV$ was
given~\cite{Abbott:1997pg,Abbott:1999ka,Abazov:2001ic}. These
results were updated at Run~II and combined with the Run~I data. The
Run~II results give $m_{D_0^1} >136 \GeV$ at 95\% confidence level
for pure $\nu\; q$ decays~\cite{Abazov:2006wp}. The combined limits
at the 95\% confidence level give $m_{D_0^1} \geq 256 \GeV$ for
${\rm Br}(D_0 \to e q) =1$ and $m_{D_0^1} \geq 234 \GeV$ for ${\rm
Br}(D_0 \to e q) =0.5$~\cite{Abazov:2004mk}. Similarly, for second
generation couplings the limits are $m_{D_0^1} \geq 251 \GeV$ for
${\rm Br}(D_0 \to \mu q) =1$ and $m_{D_0^1} \geq 204 \GeV$ for ${\rm
Br}(D_0 \to \mu q) =0.5$~\cite{Abazov:2006vc}. D0 also searched for third generation scalar leptoquarks, obtaining~\cite{Abazov:2007bs}
$m_{D_0^1} \geq 229 \GeV$ for ${\rm Br}(D_0 \to \nu_\tau b) =1$.
A similar analysis
was performed at~CDF in Run~II for scalar leptoquarks decaying to
$\ell \ell q q$ and $\ell \nu q q$ final states. For first
generation couplings the corresponding limits are $m_{D_0^1} \geq
236 \GeV,\; 205\GeV,\; 145 \GeV, \; 126\GeV$ for ${\rm Br}(D_0 \to e
q) =1,\;0.5,\;0.1,\;0.01$, respectively~\cite{Acosta:2005ge}. For
second generation couplings the limits are $m_{D_0^1} \geq 226
\GeV,\; 208\GeV,\; 143 \GeV, \; 125\GeV$ for the same branching
fractions to charged muons~\cite{Abulencia:2005et}.
Finally, D0 searched for the production of a second generation leptoquark in association
with a $\mu$. Combining with their associated production results, they obtained~\cite{Abazov:2006ej}
$m_{D_0^1} \geq 274 \GeV$ for $\lambda=1$ and ${\rm Br}(D_0 \to \mu q) =1$.

Scalar
leptoquarks can be produced on resonance at~HERA. Here the
production rates depend on the strength of the interactions
in~(\ref{WLQ}), as do the relative branching fraction to charged and
neutral leptons. Assuming equal branching fractions, the limit from
the~ZEUS experiment is $m_{D_0^1} \gappeq 290\GeV$ for $\lambda^9 =
\lambda^{10} \equiv \lambda = 0.3$ and $m_{D_0^1} \gappeq 270\GeV$
for $\lambda = 0.1$~\cite{Chekanov:2003af}. The limits from~H1 are
similar~\cite{Aktas:2005pr}.

Quasi-stable exotics require more specific search strategies which
will depend on the manner in which they interact with the elements
of the detector. The best limits come from the~CDF search for
massive charged hadrons which interact only weakly with the
calorimeter but are tracked in the muon system. The lack of observed
events puts a limit on the production cross-section for such exotic
hadrons which corresponds at the 95\% confidence level to an exotic
fermion of charge $|q| = 1/3$ of $m_D \gappeq 190
\GeV$~\cite{Acosta:2002ju}. Much weaker bounds on hadrons built from
squarks and gluinos have been obtained from ALEPH at
LEP~\cite{Heister:2003hc}.

\subsection{Indirect Bounds} \label{indirect}

There are a great many constraints on R-parity violating operators
in the MSSM\footnote{For recent reviews, see
~\cite{Chemtob:2004xr,Barbier:2004ez}. Older reviews include
~\cite{Barger:1989rk,Davidson:1993qk,Choudhury:1996ia,Allanach:1999bf,Allanach:1999ic}.}.
In addition to direct searches at colliders, there are stringent
indirect constraints from proton decay (which essentially forbid the
simultaneous presence of diquark and leptoquark operators), neutron
oscillations, $K-\bar K$ and $B- \bar B$ mixing, CP violation, rare
$B$ decays, lepton number and lepton flavor violating processes,
neutrino masses, cosmology and astrophysics, and many other sources.

Many of these processes also constrain the leptoquark and diquark
couplings $\lambda^6 -\lambda^{10}$  defined in~(\ref{WLQ})
and~(\ref{WDQ}) of the exotic supermultiplets $D$ and $D^c$. As
described in Section~\ref{E6case}, we assume that either the
leptoquark operators {\em or} the diquark operators are present, but
not both, and also that the scalar neutrinos $\nu$ and $\nu^c$ do
not acquire vacuum expectation values. In that case there are
conserved baryon and lepton numbers and R-parity, implying the
absence of proton decay (at least from the terms in~(\ref{W})) and
neutron oscillations, and also that there is no mixing between $D$
or $D^c$ and the ordinary $d$ or $d^c$ quarks. There are
nevertheless many constraints from rare processes, analogous to
those in the MSSM with R-parity violation, involving an internal $D$
and/or $D^c$ line. These were studied many years ago by Campbell et
al.~\cite{Campbell:1986xd} assuming  specific relations between the
scalar and fermion exotic masses and the other supersymmetry
breaking parameters. A reanalysis of the constraints with more
recent experimental values and general mass parameters is beyond the
scope of this paper.\footnote{Some specific processes have been
considered in~\cite{Morris:1987fm,D'Ambrosio:2001wg}.} Moreover,
there are many couplings involved when family indices are included,
often allowing individual constraints to be evaded by judiciously
tuned choices of the dominant ones, and there is also the
possibility of cancelations between diagrams. Because of these
uncertainties, we will simply utilize the most stringent MSSM
constraints for which there is a simple correspondence with a
relevant diagram for exotic particle exchange, with the
understanding that the constraints are to be considered a rough
guide rather than a rigorous limit. Furthermore, our focus is mainly
on couplings to the first two generations. Weaker bounds typically
apply to couplings to the third generation.\footnote{For that reason
the analysis in~\cite{King:2005jy} assumed that the leptoquark or
diquark couplings {\em only} involved the third generation.}

The most stringent relevant constraint on the leptoquark couplings
is from the SINDRUM II limit
\begin{equation}
\frac{\sigma (\mu^- T_i \rightarrow e^- T_i)}{\sigma (\mu^- T_i
\rightarrow \ {\rm capture})} < 4.3 \times 10^{-13}
\end{equation}
on $\mu-e$ conversion~\cite{Dohmen:1993mp}. There are several
relevant tree-level diagrams in the R-parity violating MSSM, but
only s-channel $D_0$ or $D_0^c$ exchange are relevant in the
exotic model. From the MSSM analyses~\cite{Kim:1997rr,Huitu:1997bi}
we estimate
\begin{equation} |\lambda^6_{Du^ce^c} \lambda^{6*}_{Du^c\mu^c }|< 8\times 10^{-8}
\left(\frac{m_{D_0}}{100 \ {\rm GeV}}\right)^2,
\end{equation}
or
\begin{equation}
|\lambda^6| < 3\times 10^{-4}\left(\frac{m_{D_0}}{100 \ {\rm
GeV}}\right)
\end{equation}
if we assume $|\lambda^6_{Du^ce^c}| \sim |\lambda^6_{Du^c\mu^c}| =
\lambda^6$. A similar constraint applies to $\lambda^7$ with
$m_{D_0}\rightarrow m_{D_0^c}$.

The limits on diquark couplings are much weaker. The most important
indirect limit involving the first two generations (and ignoring
possible CP-violating phases) is from the $K_L-K_S$ mass difference.
This has been analyzed in detail for the
MSSM~\cite{Barbieri:1985ty,Carlson:1995ji,deCarlos:1996yh,Slavich:2000xm,Bhattacharyya:1998be},
for which there are important contribution from box diagrams
involving four diquark vertices and also from boxes involving a $W$
exchange as well as two diquark vertices. Using the estimates
of~\cite{deCarlos:1996yh}, the most important diagrams for the
exotic case involve boxes with two internal exotic supermultiplet
lines and two internal $u$, $c$, or $t$ supermultiplet lines. Again
ignoring the possibilities of cancelations or fine-tuning the family
indices, one finds that typically
\begin{equation}
 \lambda^{9,10} < 0.04 \left(\frac{\max(m_{\tilde u_i}, m_{D_0})}{100 \ {\rm GeV}}\right)^{1/2}
\end{equation}
for the couplings involving external $d$ or $s$ and internal $u_i$.

\subsection{Cosmological Bounds} \label{cosmo}

A number of significant constraints on the properties of
quasi-stable particles, particularly quasi-stable hadrons, arise
from cosmological observations. The most severe bounds come from the
successful prediction of light element abundances by the Big Bang
Nucleosynthesis (BBN) model. When the value of the baryon-to-photon
ratio $\eta \equiv n_B/n_{\gamma}$ of $ (6.10\pm 0.21) \times
10^{-10}$ from the WMAP three year results~\cite{Spergel:2006hy} is used, the
theoretical predictions of the standard BBN scenario are in reasonable
agreement with the observed abundances of deuterium~D, $^3{\rm
He}$, $^4{\rm He}$, $^6{\rm Li}$ and $^7{\rm Li}$. These successful
predictions are based, however, on a set of assumptions about
early-universe physics. In particular they assume the Standard Model
field content and set of interactions. The presence of late-decaying
exotic particles beyond the Standard Model are therefore likely to
change the physics that ultimately gives rise to the primordial
abundances of light elements.

Constraints on long-lived exotics arising from BBN depend on two key
quantities: the abundance of the exotic and its lifetime. Shorter
lifetimes are generally less well-constrained than longer ones, but
even if the exotic particle decays wells before the epoch of BBN
(beginning roughly at the temperature scale $T\sim 0.7\; {\rm
MeV}$), there can be conflict with the successful BBN predictions if
the abundance of the exotic is too large. The relic abundance is
a function of the annihilation cross-section for
the (un-hadronized) exotics in the early universe. We will
consider the issue of relic abundance below, but first we
summarize the principal observational constraints on any new
long-lived particles.

If some new state decays with a lifetime greater than roughly a
tenth of a second, then it will decay during or after the epoch at
which BBN takes place, and the decay products
can potentially alter the predictions of
BBN. They can mediate additional interconversion between
protons and neutrons beyond that of the Standard Model interactions
at early stages of the BBN process. At later stages these decay
products can cause hadrodissociation or photodissociation of the
primordial background nuclei. The implications of these non-standard
processes are explained in detail
in~\cite{Kawasaki:2004yh,Kawasaki:2004qu}. We will here merely
sketch the conclusions before applying these results to the model of
Section~\ref{cases}.

Constraints on new long-lived particles apply to states with
lifetimes $\tau$ in the range $10^{-2}\; {\rm sec} \lappeq \tau
\lappeq 10^{12}\; {\rm sec}$. Roughly speaking, the type of
constraint depends on the epoch in which the new state decays, and
this can be divided into three temporal regions. For states decaying
with lifetime $10^{-2}\; {\rm sec} \lappeq \tau \lappeq 10^2\; {\rm
sec}$ hadronic decay products from the exotic are likely to lose
energy very quickly through electromagnetic processes. They are
therefore insufficiently energetic to destroy the newly-created
light element nuclei. Nevertheless, strong interactions allow
scattering of these decay products with protons and neutrons,
generating interconversion between the two (in particular conversion
of protons into neutrons). This is even after these baryons freeze
out with respect to electroweak conversion processes. The resulting
increase in the ratio of neutrons to protons results in a higher
yield of deuterium and $^4{\rm He}$ than is observed experimentally.

For longer-lived particles with $10^2\;{\rm sec} \lappeq \tau
\lappeq 10^7\; {\rm sec}$ the mesons from the decay products
tend to decay before they have a chance to interact with the
background neutrons and protons. Thus the $n/p$ ratio is likely to
be unchanged. But the hadronic decay products from the exotics will
be emitted with a higher kinetic energy than the thermalized nuclei
of the light elements. In this case hadrodissociation of
alpha-nuclei is common and the result is nonthermal production of
deuterium and $^6{\rm Li}$. Finally, for the case of very long-lived
exotics in which $10^7\;{\rm sec} \lappeq \tau \lappeq 10^{12}\;
{\rm sec}$ even the neutrons from the decay of the exotics will now
have time to decay before interacting with background nuclei. In
this case photodissociation from emitted photons and
hadrodissociation are competitive processes. The constraint arises
from nonthermal overproduction of $^3{\rm He}$.

\begin{figure}[tb]
\begin{center}
\includegraphics[scale=0.43,angle=0]{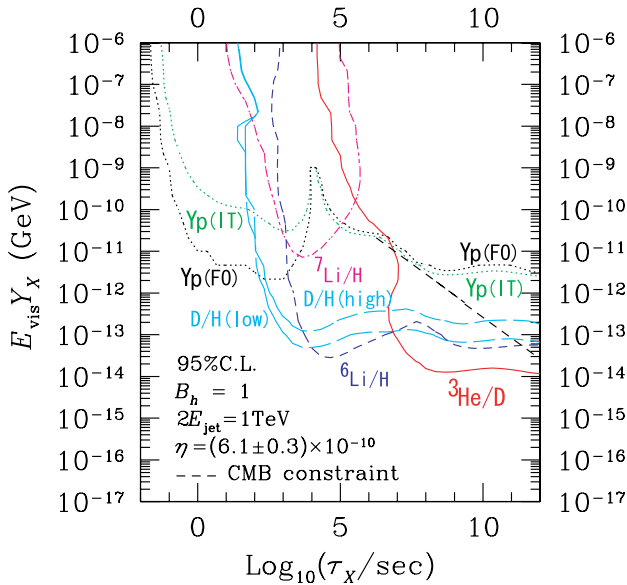}
\caption{\footnotesize \textbf{BBN Constraints on New Hadronically
Decaying Particles.} This figure, reprinted with permission from the
authors of~\cite{Kawasaki:2004qu}, summarizes the constraints
arising from BBN on a particle of mass $m_X = 1\TeV$ that decays
exclusively to hadronic Standard Model states. The various contours
represent bounds arising from the observation of different
primordial element abundances. The vertical axis indicates the relic
density $Y_X$ (i.e., the ratio of number density to entropy) of the
exotic multiplied by the amount of energy released into interacting
particles per exotic decay. For this plot that fraction is 100\%, or
$E_{\rm vis} = m_X$.} \label{fig:kaz}
\end{center}
\end{figure}

These constraints from BBN are summarized in Figure~\ref{fig:kaz},
where the region above and to the right of the various curves are
excluded.\footnote{This plot is reprinted from
Ref.~\cite{Kawasaki:2004qu} with permission of K.~Kohri.} As
indicated in the preceding paragraphs, the bounds arise from
different observations depending on the lifetime $\tau$ of the
exotic. For shorter lifetimes the principal bound arises ultimately
from observations of primordial $^4{\rm He}$ (as indicated by the
mass fraction $Y_p$) and primordial deuterium (as indicated by the
ratio D/H). The pairs of curves for these two observations reflect
different estimates for extracting the primordial abundances from
current observations. For the case of the primordial mass fraction
of $^4{\rm He}$ these estimates are from Izotov and
Thuan~(IT)~\cite{Izotov:2003xn} and from Fields and
Olive~(FO)~\cite{Fields:1998gv}. In the case of deuterium the
``high'' and ``low'' estimates differ in whether or not the recent
data of Webb et~al.~\cite{Webb:1997mt} is included in the fit. For
intermediate lifetimes the $^6{\rm Li}$ observations are most
constraining, while very long lifetimes are most constrained by
observations of $^3{\rm He}$.

The relic abundance is determined by the annihilation cross-section.
For definiteness we will assume that the lightest exotic particle
(LEP) is a fermion.\footnote{It is also quite possible that the mass
difference between the scalar and fermion is smaller than the LSP
mass. In this instance {\em both} the fermion and scalar would be
quasi-stable. Similar statements apply to heavier generations of
exotics.} The annihilation proceeds most often through QCD processes
into quarks and gluons, with a thermally-averaged cross section
\begin{equation}
\langle \sigma |v| \rangle = \frac{(14+6N_f)\pi}{27}
\left(\frac{\alpha_s^2}{m_D^2}\right) ,
\label{sigma0} \end{equation}
where the first term is for gluons and $N_f$ is the number of quark
species that may appear in the final state. For our purposes we will
take $N_f=5$. This, and the corresponding formula for scalar
LEPs,\footnote{ If the LEP is a scalar, the averaged annihilation
cross section at temperature $T$ is $\langle \sigma |v|
\rangle=\frac{(28+ 6 N_f T/m_{D_0})\pi}{27}\left(
\frac{\alpha_s^2}{m^2_{D_0}} \right)$, where the $T/m_{D_0}$ factor
is because the annihilation into quarks is $p$-wave suppressed
\cite{Jungman:1995df}. Since $T/m_{D_0}\sim 1/20$ at freezeout, the
annihilation is mainly into gluons, with a cross section about $2/3$
that for the case of a fermion LSP, and the corresponding relic
density about 50\% higher.} can easily be computed using appropriate
modifications of the cross sections for the inverse processes listed
in Appendix~\ref{appendixB}. In addition, we will assume that 50\%
of the energy is carried away by non-interacting decay products
(such as the LSP of a supersymmetric theory). With these
assumptions, the nature of the BBN constraint will depend crucially
on the exotic lifetime.

Let us consider the model suggested in section~\ref{E6case} in which
all renormalizable operators allowing for the decay of the exotic
$SU(3)$ triplets/antitriplets are forbidden. Then the first case of
allowed decay operators arise at mass dimension five via the
operators listed in~(\ref{Wdim5}). There are therefore three kinds
of decay channels for the quasi-stable fermion $D$, depending on its
mass. If the $D$ is heavier than the two superpartner bosonic
particles and/or the sum of the scalar Higgs and $S$ masses, then
three body decays into a fermion and two scalars will
dominate.\footnote{The smaller phase space for three body decays is
compensated by the larger number of channels, by color factors, and
by the $\cos^2 \beta$ suppression of the two body rates, at least
for the specific model considered here.} Should this decay be
kinematically forbidden, two-body decays into a massless fermion and
scalar Higgs or scalar $S$ may be induced through the first operator
of~(\ref{Wdim5}). As we expect our exotic states to have masses well
in excess of 100~GeV to avoid the direct search constraints of
Section~\ref{collider}, we will assume that the decay channel $D
\rightarrow d h$ is always available through this operator. Lastly,
the same operator will allow the decays $D \rightarrow d Z$, $u W$
through mixing effects. If the exotic fermion $D$ is lighter than
all scalar quarks and leptons it will decay primarily through these
processes.

For three body decays into a (massless) fermion and two (massive)
scalars the decay width is given by
\begin{equation}
\Gamma_3 =  \frac{C_i}{512\pi^3} \times \frac{m_D^3}{M_*^2} \times
K_3 , \label{Gam3}
\end{equation}
where $C_i$ is a numerical factor that depends on the particular
final state and $K_3$ represents the integral over phase space.
It is $1/3$ neglecting the scalar masses, and in general is
\begin{equation}
K_3 =  \int (2-x_A-x_B) d x_A dx_B , \label{K3}
\end{equation}
where $x_{A,B}=2E_{A,B}/m_D$ with $E_{A,B}$ the energies of scalars
$A$ and $B$. The integration limits are given in the Appendix of \cite{Barger}.
The
quantity $M_*$ is the mass scale that suppresses the
higher-dimensional operators in the superpotential. Two body decays
into a fermion and scalar are given by
\begin{equation}
\Gamma_2 = \frac{C_i m_D^3}{64\pi} \times\frac{\lang s \rang^2 \lang
h_d \rang^2}{m_D^2 M_*^2} \times { K_2^{S,V}} ,
\label{Gamma2} \end{equation}
where
\begin{eqnarray}
K_2^S &=& \(1-\frac{m_A^2}{m_D^2}\)^2 \\
K_2^V &=& \(1+\frac{2m_A^2}{m_D^2}\) \(1-\frac{m_A^2}{m_D^2}\)^2
\label{kin2} \end{eqnarray}
for decays into scalars ($m_A = m_h$ or $m_s$) and vectors ($m_A =
m_W$ or $m_Z$), respectively.

\begin{figure}[tb]
\begin{center}
\includegraphics[scale=0.45,angle=0]{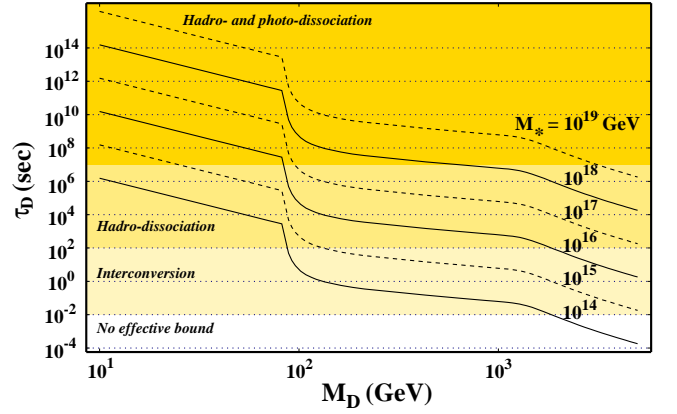}
\caption{\footnotesize \textbf{Lifetime $\tau_D$ of the Exotic Quark
in the Quasi-Stable Scenario of Section~\ref{cases}.} The lifetime
as a function of exotic mass is plotted for various values of the
suppression scale $M_*$ of $10^{19}$, $10^{18}$, $10^{17}$,
$10^{16}$, $10^{15}$ and $10^{14}$ GeV. Abrupt changes in the curves
are the result of additional decay channels opening as the mass
$M_D$ increases.} \label{fig:lifetime}
\end{center}
\end{figure}

Clearly the resulting lifetime $\tau_D$ of the exotic fermion is a
strong function of the relative masses of the states in the theory
and will be sensitive to precisely how many decay channels are
available to the exotic. To be concrete, let us consider the
$U(1)_N$ case described in Section~\ref{E6case}, with the
dimension-five operators of~(\ref{Wdim5}). Only the first operator
allows for mixing or two-body decays (assuming no vevs for the
superpartners of the Standard Model fields or $\nu^c$). A conservative estimate
of the lifetime can be made by assuming only decays into first
generation fields, allowing only those channels for which the
fermion final states are massless ({\em i.e.}, no decays into
Higgsinos or singlinos), and assuming no decays into heavy Higgs
states or $Z'$. This results in 14~partial widths for the final states: $d
Z$, $u W$, $d h$, $d s$, $d s h$, $\tilde{d}_L \tilde{u}_L u$,
$\tilde{d}_L \tilde{u}_R u$, $d \tilde{u}_L \tilde{u}_R$,
$\tilde{u}_L \tilde{e}_L \nu$, $\tilde{u}_L e \tilde{\nu}_R$, $u
\tilde{e}_L \tilde{\nu}_R$, $\tilde{d}_L \tilde{\nu}_L \nu$,
$\tilde{d}_L \nu \tilde{\nu}_R$, $d \tilde{\nu}_L \tilde{\nu}_R$.
The prefactors for these cases are given by
\begin{eqnarray}
C_{dZ}=\frac{4G_F}{\sqrt{2}} \; ; \quad C_{uW} =
\frac{8G_F}{\sqrt{2}} \nonumber \\
C_{dh}=\frac{4G_F}{\sqrt{2}}\frac{\sin^2\alpha}{\cos^2\beta} \; ; \quad
C_{ds}= \frac{1}{\lang s \rang^2} \nonumber \\
C_{dsh}=\sin^2\alpha/4 \; ; \quad C_{q \tilde{q} \tilde{q}} =4\; ,
\label{Cfac} \end{eqnarray}
and unity for the final states arising from the final operator
of~(\ref{Wdim5}). $\alpha$ is the usual MSSM mixing angle for the Higgs scalar mass eigenstates.
For the decays involving the second operator
in~(\ref{Wdim5}), with only first generation particles, the two
fields arising from the two quark doublets (of which there are two
possible combinations) must not have the same color. Hence the total
counting factor gives $C_i = 4$.

For completeness, we also consider the decays mediated by virtual
$W$'s and $Z$'s for $m_D \lappeq m_W$ (decays involving virtual $h$,
$s$, or $Z'$ are much less important than other allowed decays for
all masses). The result of a straightforward calculation, using the
partial rates calculated in~\cite{Barger:1985nq}, is
\begin{equation}
\Gamma_{W^*+Z^*}= \theta^2 \frac{G_F^2 m_D^5}{192 \pi^3}\frac{
[243-162\sin^2 \theta_W + 242 \sin^4 \theta_W]}{18},
\label{virtual}
\end{equation}
where
\begin{equation} \theta=\frac{\lang s\rang \lang h_d \rang}{M_* m_D}
\end{equation}
is the $D_L-d_L$ mixing angle (assumed small) and $\sin^2 \theta_W
\sim 0.23$ is the weak angle. In~(\ref{virtual}) we have assumed
that the $D$ mixes only with the $d$, neglected CKM mixing, and
included decays into three families of massless leptons and five
flavors of massless quarks. $W-Z$ interference and identical
particle effects are included for the $d u \bar u$ and $d d \bar d$
channels, respectively.

To obtain numerical estimates it is necessary to postulate specific
mass values for certain scalar fields. Let us take $m_{\tilde{f}} =
500 \GeV$ for all scalars of the MSSM, $m_s = 1000 \GeV$ for the
singlet scalar field,\footnote{We will only consider the real part
of this scalar field to be dynamical. When the
models of this paper are embedded in theories with additional
$U(1)$'s it is the imaginary part of this field that is typically
``eaten'' to produce a massive $Z'$ boson.} $m_h = 115 \GeV$, $\lang
s \rang = 740\GeV$ and $\tan\beta=10$. We also will take the
decoupling limit for the MSSM Higgs sector such that the mixing
angle $\alpha$ is given by $\alpha = \beta - \pi/2$. The resulting
lifetime is plotted in Figure~\ref{fig:lifetime} as a function of
exotic fermion mass $M_D$ and the mass scale of the dimension five
operator $M_*$. As the actual numerical values of these results are
sensitive to the masses of the scalars (as well as the number of
decay channels considered), the results in Figure~\ref{fig:lifetime}
should be taken as indicative of what is likely in models of this
sort. For values of $M_D \leq m_h$ decays can only proceed via
mixing with Standard Model fermions; thus the lifetime decreases
slowly with increasing $M_D$. Most of this range can generally be
excluded from direct collider searches (though this is a
model-dependent statement) as discussed in Section~\ref{collider}.
Above this mass scale the lifetime generally decreases rapidly,
particularly for $M_D$ greater than twice the typical scalar mass of
the squarks and sleptons. We conclude from Figure~\ref{fig:lifetime}
that for reasonable exotic masses and suppression scales $M_*$ (i.e.,
$M_*$ less than or equal to the reduced Planck mass) the primary
constraints will involve proton-neutron conversion and
hadrodissociation of background alpha nuclei.

\begin{figure}[tb]
\begin{center}
\includegraphics[scale=0.41,angle=0]{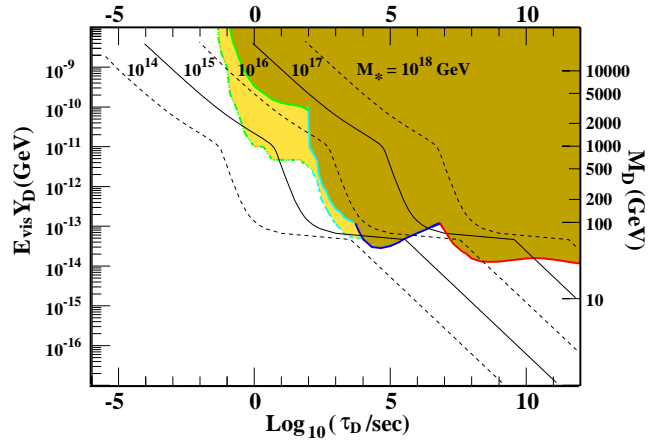}
\caption{\footnotesize \textbf{BBN Constraints on Exotic Quark
Parameters.} We plot combinations of lifetime $\tau_D$ and $M_D$ (or
equivalently $E_{\rm vis} Y_D$ for $M_*$ of  $10^{18}$,
$10^{17}$, $10^{16}$, $10^{15}$ and $10^{14}$ GeV. The darker shaded region is
the union of all constraints from Figure~\ref{fig:kaz} using the
weakest bounds for D/H and $Y_p$, while the lighter shaded region
extends this disallowed space by using the more restrictive values
for these quantities. Note that this plot assumes that 50\% of the
mass energy of the exotic is carried away by non-interacting LSPs
(such that $E_{\rm vis} = 0.5 M_D$).} \label{fig:qscosmo}
\end{center}
\end{figure}

From the cross-section~(\ref{sigma0}) the freeze-out temperature
$T_f$ and relic density $Y_D$ of the exotic particle can be computed
using standard techniques~\cite{Jungman:1995df}. For each value of
$M_*$ there exists a unique pair of values for $\tau_D$ and $E_{\rm
vis} Y_D$ (or $\tau_D$ and $M_D$) which can be compared with the
experimental bounds. In Figure~\ref{fig:qscosmo} we have plotted
these contours as a function of $M_*$ for the values
$10^{18}$, $10^{17}$, $10^{16}$, $10^{15}$ and $10^{14}$ GeV and overlaid them
on the limits from Figure~\ref{fig:kaz}. The darker shaded region is
the union of all constraints from Figure~\ref{fig:kaz} using the
weakest bounds for D/H and $Y_p$, while the lighter shaded region
extends this disallowed space by using the more restrictive values
for these quantities. Along each curve the value of $M_D$ increases
from very small to very large values as one moves from the lower
right to the upper left. For large suppression factors ($M_* \gappeq
10^{17} \GeV$) only extremely light exotics (or extremely massive
ones) can be tolerated by existing limits. Such small values would
likely be in conflict with collider bounds from the Tevatron. For
$M_* \lappeq 10^{15}\GeV$ nearly the entire range of mass values
$M_D$ is allowed by current observations.

We again stress that there are several factors that can change the
impact of the light element abundance constraints on the exotic
mass. Some of these factors would serve to weaken the bounds on
$M_D$ for a given $M_*$: changing the scalar masses for
superpartners (in particular $m_s$), allowing additional decay
channels into Higgsinos, singlinos and second/third generation
quarks, allowing additional energy to be carried off by
non-interacting particles,
and so
forth. Furthermore, it has recently been suggested that hadronized
exotics should be expected to undergo a second round of annihilation
shortly after the QCD phase transition~\cite{Arvanitaki:2005fa}.
This can serve to dilute these bounds considerably as well. We
therefore conclude that it is not unreasonable to assume that should
string-inspired exotic states exist, with decay amplitudes
suppressed by a mass scale somewhat below the reduced Planck scale, then
they may very well exist in mass ranges relevant to the upcoming LHC
experiment. Such scales could emerge in string constructions with
some dimensions larger than the Planck length, or be associated with non-Planck scale physics.
In the next section we will consider the
phenomenological implications of this scenario, before returning to
the case of prompt decays in Section~\ref{decay}.

\section{Quasi-Stable Exotics at the LHC} \label{qs} 

Given the discussion from the previous section, it is reasonable to
assume that exotic new triplet/anti-triplet $SU(3)$ representations
will hadronize into color-singlet states on a rapid time scale.
These exotic hadrons have come to be known as {\em R-hadrons} in the
literature and we will adopt that name for them here. These
R-hadrons will be stable on the time scale of the detector and can
be treated as hadrons for the duration of their interaction with the
detector elements. There has been some discussion of such
quasi-stable exotics in the literature, but this has been primarily
with regard to long-lived gluino-based
R-hadrons~\cite{Baer:1998pg,Mafi:1999dg,Kraan:2004tz,Kraan:2005ji,Kilian:2004uj,Hewett:2004nw}.
While the phenomenology of long-lived triplet/anti-triplet
representations will be similar, there are important differences. We
will therefore revisit the subject in this section and focus on the
signature of such states at the LHC experiments. Our analysis will
be sufficient to give a sense of the discovery reach of the LHC for
such new states, though a more refined analysis with a full detector
simulation will no doubt sharpen the conclusions obtained here.

\subsection{Collider Phenomenology}
\label{qspheno}

Unlike the previous studies which assume a gluino component for the
R-hadron, here the exotic component can be a scalar or a fermion.
Furthermore, these earlier papers were generally motivated by
scenarios with very heavy scalars, thus assuming only direct pair
production of gluinos. Here there is the very real possibility of
producing pairs of the heavier states in the supermultiplet; the
next-to-lightest exotic particles (NLEPs). Depending on the mass
differences involved, these states may then decay into the LEP which
subsequently hadronizes. This provides an additional handle for
triggering and event reconstruction which we will describe in the
next subsection.

Another important distinction between the case of the gluino and
that of the exotic quark is in their $SU(3)$ representation, which
impacts the sorts of R-hadrons that can form and their interaction
cross-sections with the detector elements. Like their gluino
counterparts, exotic triplets and anti-triplets can form R-mesons
and R-baryons, with the former kinematically favored at the initial
hadronization stage. However, the R-hadrons considered here will
necessarily have one fewer ``active'' quark than those of the
gluino-based variety. As a result, the total cross-section for
R-hadron interactions with protons and neutrons will be
proportionally reduced.

The R-meson states will include $D\bar{d}$ and $D\bar{u}$
combinations (as well as the anti-states) and the R-baryons will
include $Ddd$, $Duu$ and two combinations of $Ddu$ (as well as their
anti-states).\footnote{Approximately 15\% of the R-hadrons produced
in the primary high-$p_T$ process will involve a strange quark, with
a resulting $6\;{\rm mb}$ reduction in their interactions with
protons and neutrons. As this will have a negligible impact on the
collider physics considered below we will neglect all heavy flavor
R-hadrons.} The meson states will be approximately degenerate in
mass, with the bulk of the mass being accounted for by the exotic
component. We thus expect roughly 50\% of the R-mesons formed at the
initial vertex to be charged. Once produced, charged R-mesons will
leave tracks in the inner detector elements. The R-mesons will pass
largely unaffected through the electromagnetic calorimeter and enter
the hadronic calorimeter. Here the R-hadrons will undergo a series
of interactions which include elastic scattering off nucleons,
charge-exchange interactions with nucleons and
meson-to-baryon/baryon-to-meson interactions~\cite{Kraan:2004tz}.
These processes yield an interaction cross-section of roughly 12~mb
for R-mesons and 24~mb for R-baryons. The array of possible
interactions depends on the R-hadron itself and reveals an important
asymmetry.

R-mesons of the form $D\bar{q}$ will preferentially undergo
meson-to-baryon transitions by extracting two quarks from a target
nucleon and producing a light pion in the final state. The resulting
$Dqq$ R-baryon will {\em remain} an R-baryon due to the absence of
anti-quarks in the detector material. Among the possible R-baryons
which can form, the case $Dud$ with $d$ and $u$ in an s-wave
configuration will be the lightest state, with the p-wave
configurations of $Dud$, $Ddd$ and $Duu$ more massive. In contrast,
R-mesons of the form $D^c q$ will {\em remain} R-mesons due to the
lack of quark-antiquark annihilation possibilities. Even if an
R-baryon of the form $D^c \bar{q} \bar{q}$ were to form, it would
quickly be destroyed by baryon-to-meson interactions which produce
light pions.

\begin{figure}[tb]
\begin{center}
\includegraphics[scale=0.5,angle=0]{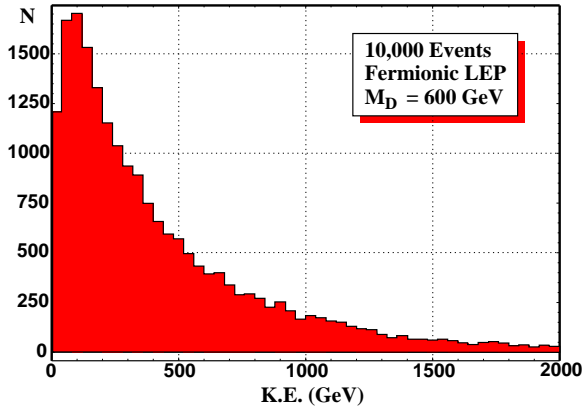}
\caption{\footnotesize \textbf{Initial Kinetic Energy of Produced
R-hadrons.} Distribution of kinetic energies for fermionic LEPs with
$m_D = 600 \GeV$ prior to entering the calorimeter system.}
\label{fig:keLEP}
\end{center}
\end{figure}

With the cross-sections given above, we expect the exotic R-hadrons
to undergo a number of interactions with the hadron calorimeter. For
R-baryons (R-mesons) we estimate on average 8~(6)~interactions in
the ``barrel'' region ($0\leq \eta \leq 1.5$) and
10~(7)~interactions in the ``endcap'' region ($1.5 \leq \eta \leq
2.4$) for both the ATLAS and CMS detectors prior to entering the
muon system. As the bulk of the energy and momentum is carried by
the (non-interacting) exotic quark, these interactions tend to
result in a very small energy deposit in the calorimeter
cells~\cite{Kraan:2004tz}. The exact amount depends on the mass of
the R-hadron, its kinetic energy and the material through which it
is passing. As an example, consider the pair production via~QCD of
exotic fermion LEPs with mass $M_D = 600 \GeV$.
Figure~\ref{fig:keLEP} gives the distribution of kinetic energies of
the produced LEPs from a simulation 10,000~events (roughly $10 {\rm
fb}^{-1}$ of data-taking). While most are produced with relatively
little kinetic energy there is a long tail in the distribution, with
a mean at $E_{\rm kin} \simeq 400 \GeV$. From the results
of~\cite{Kraan:2005ji} we estimate the typical energy loss per
interaction for R-hadrons with $E_{\rm kin} \lappeq 1 \TeV$ to be
somewhere between $0.2$ and $2.2$ GeV. For our simulations in the
next subsection we will assume $1.5$ GeV in energy loss per nuclear
interaction.

We therefore expect most, but not all, R-mesons produced in the
primary interaction with $|\eta|\leq 2.4$ to punch-through to the
muon chambers as some form of R-hadron in the ATLAS or CMS detector.
Those including $D$ will most often arrive as neutral R-baryons
while those involving $D^c$ will arrive as some form of neutral or
charged R-meson. If we neglect the small possibility of $D^c s$
R-mesons, we would anticipate roughly 50\% of the R-mesons involving
$D^c$ to be charged when they enter the muon chamber. These charged
R-hadrons will leave tracks in the muon system and will have a
characteristic velocity $\beta$ significantly different from the
essentially massless muon. Of course it is possible for R-hadrons of
all types to further interact hadronically in the muon system --
perhaps in a charge-exchanging way. We will neglect that small
possibility in what follows.

Typical distances between the widely separated resistive plate
chambers in the muon systems of ATLAS and CMS are less than 1~meter,
with a separation between the first and last such plate at
approximately 3~meters~\cite{TDR}. It should therefore be possible
to make a robust distinction between tracks arising from the exotics
and those from background muons by measuring the respective
times-of-flight (TOF). The primary source of background muons will
be single and double weak boson production. Separating the signal
from background can be performed by requiring that the TOF between
any pair of reference points for the candidate R-hadron be at least
3~ns greater than the TOF of a muon (with $\beta \simeq 1$). The
value of 3~ns is sufficiently greater than the $\delta t$ resolution
of both the ATLAS and CMS muon system to provide a highly
significant $S/\sqrt{B}$ value~\cite{Kraan:2005ji}. The final
requirement is that the R-hadron be moving sufficiently {\em
swiftly} to arrive at the muon chamber with the time-window for the
data to be recorded with the current bunch crossing. This
corresponds to a TOF of approximately 18~ns to reach the muon
system, or $\beta \simeq 0.5$ for the R-hadron~\cite{Nisati:1997gb}.

\subsection{Discovery Reach}

There remains, however, the question of triggering on these events.
What fraction of the R-hadron events will eventually be recorded to
tape? The low-level trigger must capture an event prior to full
reconstruction, thus the presence of tracks in the inner detector
will not be sufficient to capture an event without significant
activity in either the calorimeters or the muon chambers. R-hadrons
which stop in the calorimeter will likely deposit sufficient energy
to trigger in the $E_T^{\rm sum}$ channel or $E_T^{\rm miss}$
channel (assuming the other R-hadron punches-through or decays at a
different time). However, given the potentially very long lifetimes
of these states these decays will occur at a much later
bunch-crossing than that which produced the exotic quarks. Such
issues have been addressed in the context of long-lived
gluinos~\cite{Arvanitaki:2005nq}. We will not pursue the
phenomenology of these cases further, though we will note the number
of such ``stopping'' exotics in our simulations to follow.

The punch-through R-hadrons are likely to leave only 10-50 GeV of
transverse energy in the detector -- not enough to trigger in the
$E_T^{\rm sum}$ channel. Furthermore, as the exotic quarks are pair
produced and are back-to-back in the center-of-mass frame, the total
amount of {\em transverse} energy carried away by the exotics is
small. Of course, production of the next-to-lightest exotic particle
(NLEP) allows for the possibility of either increased $E_T^{\rm
miss}$ through $D_0 \to D_{1/2} \wtd{\chi}^0_1$ decays or increased
jet activity through $D_0 \to D_{1/2} \wtd{g}$ decays. In the former
case, the two $D_{1/2} \wtd{\chi}^0_1$ systems are again
back-to-back, with only a slight increase in $E_T^{\rm miss}$;
typical values are less than $50 \GeV$.

If either of these triggers is utilized, the event will almost
certainly pass the second level triggers as well (due to the
presence of, say, a charged track leading to the calorimeter cells
and/or the presence of a track in the muon chamber). Nevertheless,
it is still preferable for a track to be identified in the muon
chamber for signal extraction and correct particle identification.
We thus consider the muon trigger alone. Since we expect the
emergence of only one charged R-meson (at most) from the two
produced exotics, the appropriate low-level trigger is the
single-muon channel. We will require a very conservative $p_T$
threshold of 15~GeV for triggering on the R-mesons that enter the
muon chamber (after accounting for the energy loss due to hadronic
interactions in the calorimeter).

\begin{table}
{\begin{center}
\begin{tabular}{|c||c|c|c|c|c|}
\multicolumn{1}{c}{} & \multicolumn{5}{c}{Benchmark Point} \\
\hline
 & \parbox{1.0cm}{A} &
\parbox{1.0cm}{B} & \parbox{1.0cm}{C} &
\parbox{1.0cm}{D} & \parbox{1.0cm}{E} \\
\hline \hline
Geom. Accept. & 75.5\% & 79.9\% & 82.3\% & 86.8\% & 82.5\% \\
Charged Frac. & 25.2\% & 25.0\% & 25.1\% & 25.2\% & 25.4\% \\
Temp. Accept. & 82.7\% & 82.8\% & 81.9\% & 79.1\% & 76.9\% \\
TOF & 97.3\% & 96.5\% & 97.2\% & 97.3\% & 97.0\% \\ \hline
Total Accept. & 15.3\% & 16.0\% & 16.5\% & 16.9\% & 15.6\% \\ \hline
\hline
$N_{\rm signal}$ ($\times 10^3$) & 120 & 119 & 119 & 11.2 & 26.6 \\
$N_{\rm stop}$ ($\times 10^3$) & 11.1 & 10.8 & 11.3 & 1.36 & 4.56
\\ \hline
%
%
\end{tabular}
\end{center}}
{\caption{\label{qscases} {\bf Signal Acceptance for Quasi-Stable
R-hadron Scenarios}. Geometrical acceptance represents the fraction
of R-hadrons that are produced with $|\eta| \leq 2.4$. Temporal
acceptance represents the fraction of charged non-stopping R-hadrons
that arrive within 18~ns of the primary interaction for the event.
The percentage that traverse a 3~meter fiducial distance at least
3~ns slower than a $\beta=1$ muon would is given by TOF. The product
of these fractions is the total acceptance. The number of signal
events (as well as the number of stopping R-hadrons) is given for
$10\ {\rm fb}^{-1}$ of integrated luminosity.}}
\end{table}

\begin{figure}[tb]
\begin{center}
\includegraphics[scale=0.5,angle=0]{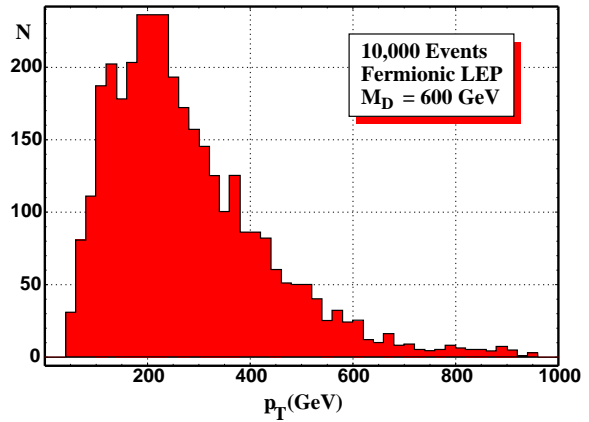}
\caption{\footnotesize \textbf{Transverse Momentum of Charged
R-mesons.} The distribution of $p_T$ for charged R-Mesons with
$\beta \geq 0.5$ upon entering the muon system. We assume a minimum
of $p_T \geq 15\GeV$ to trigger on the charged track. All R-hadrons
moving with the minimum velocity have sufficient momentum to meet
this threshold.} \label{fig:ptLEP}
\end{center}
\end{figure}

We illustrate the effectiveness of this search strategy in
Table~\ref{qscases} for the five benchmark cases presented in
Section~\ref{masses}. The acceptance at each stage is roughly
constant across the benchmark scenarios. Approximately~75-85\% of
the R-hadrons are produced with $|\eta| \leq 2.4$, and roughly 25\%
emerge from the calorimeter into the muon system as charged mesons.
Of these, approximately 80\% have $\beta \geq 0.5$ and thus arrive
within 18~ns of the primary interaction in the event. Each of these
R-hadrons will therefore produce a charged track in the muon system.
The distribution in transverse momentum for these objects upon
arrival at the muon chambers is given in Figure~\ref{fig:ptLEP} for
Scenario~C. In this case all of the R-mesons have sufficient $p_T$
to trigger given our 15~GeV minimum $p_T$ requirement. This fact was
true of all five benchmark points. Thus adding additional trigger
possibilities (such as $E_T^{\rm miss}$) is unlikely to add
significant numbers of signal events if the muon system is to be
used for particle identification. Finally, the fraction of R-mesons
moving sufficiently slowly to traverse a 3~meter fiducial distance
at least 3~ns longer than a $\beta =1$ muon is given by the ``TOF''
entry in Table~\ref{qscases}. This represents the vast majority of
R-mesons that enter the muon system within the 18~ns time window. We
therefore estimate the total acceptance to be approximately
one-sixth of all produced quasi-stable exotics.

The discovery reach will track the production cross-section for the
lightest exotic particle. In each of our benchmark cases there is at
least one exotic state with a mass below 450 GeV, providing for
copious production at LHC energies. For the simulations described in
Table~\ref{qscases} we allow for production of both the exotic
fermion and the lighter exotic scalar. In case~A both the fermion
LEP and its slightly heavier scalar will be quasi-stable, while the
other cases will involve SUSY cascade decays for the heavier states.
The higher production cross-section for the fermion (as demonstrated
in Section~\ref{production}) reflects itself in the factor of ten
between the number of signal events that arise for cases~A-C and
those of cases~D and~E. Given the exceptionally large
signal-to-background ratio for events of this type, discovery will
not prove a problem if such light exotics exist. We also note that a
fair number of the produced exotics have insufficient kinetic energy
to punch-through to the muon system and will stop in the
calorimeter. The issue of detecting these events has been addressed
elsewhere~\cite{Arvanitaki:2005fa}. As the lifetime of these exotics
is an undetermined parameter, we merely list the number of such
events in Table~\ref{qscases}.

\begin{figure}[tb]
\begin{center}
\includegraphics[scale=0.5,angle=0]{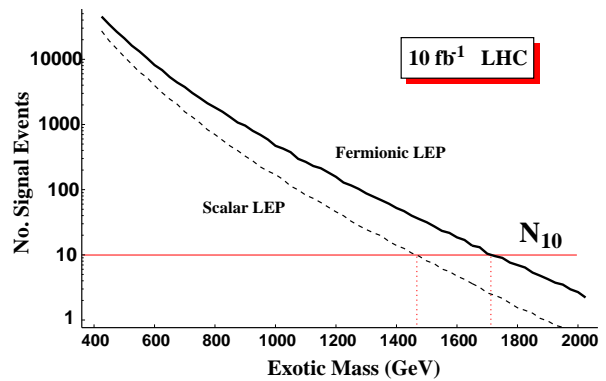}
\caption{\footnotesize \textbf{Reach in Exotic Mass for Muon
Signature.} Discovery reach at the ATLAS experiment for
``punch-through'' quasi-stable exotics. The threshold for discovery
is taken to be ten muon-like events, assuming a negligible
background rate.} \label{fig:reach}
\end{center}
\end{figure}

\begin{figure}[tb]
\begin{center}
\includegraphics[scale=0.5,angle=0]{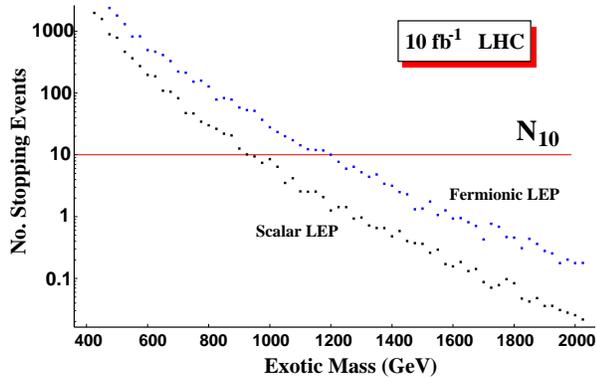}
\caption{\footnotesize \textbf{Stopping R-hadrons as a Function of
Exotic Mass.} Discovery reach at the ATLAS experiment for
quasi-stable exotics which stop in the calorimeter and subsequently
decay. The threshold for discovery is taken to be ten such events,
assuming a negligible background rate.} \label{fig:stop}
\end{center}
\end{figure}

As the masses of the exotics increase, the production rate falls. We
estimate the discovery reach in $m_D$ and $m_{D_0}$ via the muon
signature channel in Figure~\ref{fig:reach}. We use the standard
practice of taking ten events to constitute the discovery threshold
in cases with an exceptionally small number of expected background
events. This corresponds to a mass reach of $m_D \lappeq 1700 \GeV$
for the fermion and $m_{D_0} \lappeq 1450 \GeV$ for the scalar
exotic in 10 fb$^{-1}$ of integrated luminosity. Since it
should also be able to detect the fraction of events that stop
somewhere in the detector itself, we also include the mass reach for
this signature in Figure~\ref{fig:stop}. Here the mass reach for 10
events is $m_D \lappeq 1100 \GeV$ for the fermion and $m_{D_0}
\lappeq 925 \GeV$ for the scalar. These rough estimates for the
discovery reach of quasi-stable $SU(3)$-charged objects are in good
agreement with calculations performed in other model
contexts~\cite{Kraan:2005ji,Kilian:2004uj,Hewett:2004nw}.


\section{Prompt Decay Signatures} \label{decay} 

Our intention is to study the presence of exotic supermultiplets in
the presence of ``standard'' supersymmetric states such as those of
the MSSM. This is motivated by a number of considerations. First,
the study of the collider signatures of $E_6$-inspired exotics in
isolation has been performed by a number of authors. The signatures
and analysis strategies for these cases are by now well understood.
For example, a scalar state which decays directly into a lepton +
quark or into two quarks can be reliably reconstructed by looking at
the invariant mass distribution of leptons + jets and pairs of jets,
respectively. This is possible even in the presence of Standard
Model backgrounds and a measurement of the mass of the exotic scalar
can readily be made~\cite{Cakir:2005iw}. It is therefore more
interesting to consider the situation when a richer spectrum of
``new physics'' is present.

Second, most theoretical models of supersymmetric physics that have
an ultraviolet completion -- whether a GUT model or some sort of
string construction -- tend to predict that the exotics described in
Section~\ref{cases} will be present {\em in conjunction} with the
states of the MSSM. In fact, as alluded to in the Introduction, they
will typically arise with additional gauge fields (and their gaugino
partners) and additional scalars. Therefore, considering the exotics
in isolation may oversimplify the experimental challenges associated
with these fields. Finally, it is interesting to ask how the
presence of MSSM superpartners can serve as ``background'' to the
identification, extraction and eventual study of exotic new
supermultiplets -- and indeed how the exotic states can complicate
otherwise well-established analysis routines for supersymmetric
parameter measurements.

In this section we will study the collider signatures of our exotic
benchmark cases from Table~\ref{benchmarks}. We will focus primarily
on the leptoquark scenario for technical reasons, but we will
occasionally address the diquark scenario as well. We reserve a full
treatment of the diquark cases to a future publication. We allow the
full set of interactions between exotics and MSSM states given in
equations~(\ref{WLQ}) or~(\ref{WDQ}), but restrict ourselves to
interactions between the exotic states and the first generation of
Standard Model/MSSM states. We will also restrict the interactions
with gauginos and Higgsinos to include only those states of the MSSM
(i.e., we continue to neglect the possible presence of additional
singlets or $U(1)$ factors). This truncation reduces the
model-dependence of the study considerably and corresponds to
plausible limits of full-scale extended
models~\cite{Barger:2005hb,Barger:2006kt}. We will also restrict
attention to a single exotic fermion and the lighter exotic scalar,
and will ignore the possibility of heavier generations of exotic
particles. The latter would generally be expected to decay rapidly
to the lightest generation if they mix and are sufficiently heavy,
or else to exhibit production and decay patterns similar to to those
of the lightest generation. Finally, we ignore decays into the right
handed neutrino $\nu^c$ and its scalar partner. Depending on the
model of neutrino mass generation, their masses might or might not
be relevant at the LHC~\cite{Kang:2004ix}. If they are sufficiently
light, the effects of the $\lambda^8$ term in (\ref{WLQ}) would be
similar to those from the neutrino part of the $\lambda^7$ term.

\subsection{Decays of Exotics}
\label{brats}

The majority of previous studies on the collider phenomenology of
leptoquarks and diquarks have focused on non-supersymmetric
scenarios in which the exotic state decays directly to Standard
Model states. This may arise through mixing operators or by
considering only the production and decay of scalar exotics with
branching fractions to pairs of SM fermions set to
unity~\cite{Andre:2003wc,Mehdiyev:2006tz,Hewett:1987yg,Blumlein:1996qp,Dion:1997jw,Eboli:1997fb,Dion:1998wr}.
In order to discuss the full range of supersymmetric decays
available -- and to study the interesting issues outlined in the
introduction to this section, it is necessary to postulate a fixed
spectrum of MSSM states to include with the five scenarios of
Table~\ref{benchmarks}.

\begin{table}[thb]
{\begin{center}
\begin{tabular}{|l|c||l|c|} \hline
Param. & SPS~1a & Param. & SPS~1a \\ \hline
$m_{\wtd{\chi}_{1}^{0}}$ & 99.9 & $m_{\tilde{t}_{1}}$ & 381.4 \\
$m_{\wtd{\chi}_{2}^{0}}$ & 188.4 & $m_{\tilde{t}_{2}}$ & 587.3 \\
$m_{\wtd{\chi}_{3}^{0}}$ & 375.5 & $m_{\tilde{c}_{1}}$, $m_{\tilde{u}_{1}}$ & 535.3 \\
$m_{\wtd{\chi}_{4}^{0}}$ & 394.0 & $m_{\tilde{c}_{2}}$, $m_{\tilde{u}_{2}}$ & 554.5 \\
%
$m_{\wtd{\chi}_{1}^{\pm}}$ & 187.7 & $m_{\tilde{b}_{1}}$ & 504.5 \\
$m_{\wtd{\chi}_{2}^{\pm}}$ & 394.7 & $m_{\tilde{b}_{2}}$ & 535.0 \\
$m_{\tilde{g}}$ & 627.9 & $m_{\tilde{s}_{1}}$, $m_{\tilde{d}_{1}}$ & 534.4 \\
B-ino\% & 97.4\% & $m_{\tilde{s}_{2}}$, $m_{\tilde{d}_{2}}$ & 559.3 \\
%
%
$m_{h}$ &  111.7 & $m_{\tilde{\tau}_{1}}$ & 145.5 \\
$m_{A}$ & 412.7 & $m_{\tilde{\tau}_{2}}$ & 220.6 \\
$m_{H}^{\pm}$ & 420.3 & $m_{\tilde{\mu}_{1}}$, $m_{\tilde{e}_{1}}$ & 145.8 \\
$\mu$ & 369.4 & $m_{\tilde{\mu}_{2}}$, $m_{\tilde{e}_{2}}$ & 211.4
\\ \hline
\end{tabular}
\end{center}}
{\caption{\label{spectraSUSY}\footnotesize {\bf Sample non-exotic
spectra for Snowmass point~1a (SPS~1a)}. All masses are in GeV. }}
\end{table}

Our MSSM ``background'' sample corresponds to the minimal
supergravity point SPS~1a, from the Snowmass Points \&
Slopes~\cite{Allanach:2002nj}. This point has a spectrum determined
from the following set of minimal supergravity parameters: a unified
scalar mass $m_0 = 100 \GeV$, a unified gaugino mass $m_{1/2} = 250
\GeV$, a unified trilinear scalar coupling $A_0 = -100 \GeV$,
$\tan\beta = 10$ and positive $\mu$ parameter. These soft
supersymmetry-breaking values are specified at a high-energy input
scale, here assumed to be the GUT scale ($\Lambda _{\GUT} = 1 \times
10^{16} \GeV$). These GUT-scale parameters are passed directly to
{\tt PYTHIA}, which then generates the physical mass spectrum for
the MSSM at the electroweak scale. This process involves using
approximate analytical solutions to the renormalization group
equations (RGEs) for the soft Lagrangian parameters. Superpartner
masses are computed at tree level, except for the Higgs sector which
is computed from the full one-loop effective potential. The result
is given in Table~\ref{spectraSUSY}. A more careful treatment of the
RGEs and/or the physical eigenstate masses might correct these
numbers by as much as 10\%, but the accuracy achieved here is
sufficient for out purposes.

Snowmass Point~1a is the benchmark point from~\cite{Allanach:2002nj}
most favorable for SUSY discovery at the LHC. It is therefore one of
the most-studied supersymmetric models in the literature. All the
superpartners are within reach of the LHC (and indeed, some may be
detectable at the Tevatron with sufficient integrated luminosity).
The relatively light scalar leptons are particularly important for
this benchmark point, as they significantly enhance the event rate
for trileptons and large missing transverse energy ($\met$) through
the production of electroweak gauginos. The relatively light gluino
also ensures a large rate for multi-jet events with large $\met$. We
will discuss these properties in the following subsections.

\begin{figure}[t]
\begin{center}
\includegraphics[scale=0.6,angle=0]{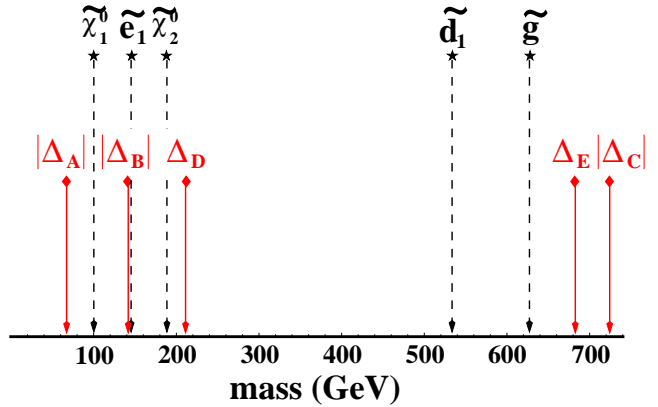}
\caption{\footnotesize \textbf{Relative mass scales for SPS~1a and
exotic cases of Table~\ref{benchmarks}.} The mass difference between
the exotic fermion and the lightest scalar, given by the quantity $\Delta_1$ in~(\ref{delta1})
is given for each of the five cases summarized in
Table~\ref{benchmarks}. We indicate with dotted arrows the masses of
some of the relevant MSSM states that may appear in the decay
chains.} \label{fig:spectrum}
\end{center}
\end{figure}

The five cases in Table~\ref{benchmarks} were selected to provide a
range of possible decay chains between the exotic states, the MSSM
spectrum and Standard Model states. The most important of the
relevant mass scales are shown schematically in
Figure~\ref{fig:spectrum}. Mass differences between the exotic fermion and the lightest
exotic scalar are given in terms of the
$\Delta_1$ variable introduced in~(\ref{delta1}) for the five cases
(note that $\Delta_1 < 0$ for cases~A-C). Key thresholds are
indicated by superpartner masses for the two lightest neutralinos
$\wtd{\chi}_{1,2}^{0}$, the gluino $\wtd{g}$, the lightest
first-generation sleptons $\wtd{e}_1$ and the lightest
first-generation squarks $\wtd{d}_1$. Note that the exotic fermion
and its lightest scalar partner are each above 300~GeV in mass in
all five cases.

In the first three cases of Table~\ref{benchmarks} the fermion is
the lightest exotic particle (LEP). For the leptoquark scenario the
two-body decay channels to a scalar lepton and a (massless) quark
are always available due to the very low-mass sleptons in the SPS~1a
model. Taking $\lambda^6 = \lambda^7 = 0.1$, the branching fractions
for $\LQ_{1/2}$ into the final states $d \; \wtd{\nu}_e^L$, $u \;
\wtd{e}_R^-$ and $u \; \wtd{e}_L^-$ are 28\%, 50\% and 22\%,
respectively for cases~A-C. For the diquark scenario the equivalent
two-body decays to a quark and squark are kinematically forbidden.
In this case the fermionic diquark must decay via three-body
channels into states involving gauginos and two SM quarks. For our
analysis we will only consider cases in which the final state
involves the LSP neutralino. These three-body decay rates are
given in Appendix~B for decays via first-generation virtual squarks,
as are those for the leptoquark case (relevant  in different kinematic regimes
for which the two-body decays are forbidden).

\begin{figure}[t]
\begin{center}
\includegraphics[scale=0.48,angle=0]{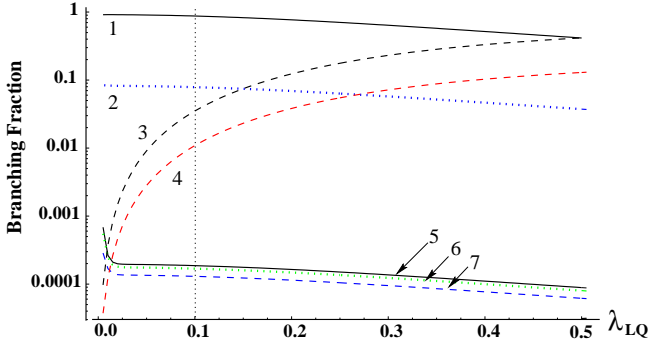}
\caption{\footnotesize \textbf{Branching ratios into various final
states for $\LQ_{0}$ decays in Scenario~C.} The branching fraction
into the seven allowed final states for the scalar leptoquark are
given as a function of a (universal) Yukawa coupling $\lambda^6 =
\lambda^7 \equiv \lambda_{\rm LQ}$. These allowed final states are
as follows: (1) $D_{1/2}\,\tilde{g}$ (2) $D_{1/2}\,\tilde{\chi}_1^0$
(3) $u\,e$ (4) $d \, \nu_e$ (5) $D_{1/2}\,\tilde{\chi}_2^0$ (6)
$D_{1/2}\,\tilde{\chi}_3^0$ (7) $D_{1/2}\,\tilde{\chi}_4^0$. }
\label{fig:Cbrats}
\end{center}
\end{figure}

For both the leptoquark and diquark scenarios the direct decay into
Standard Model fermions is the only channel kinematically available
to the exotic scalar in Case~A. The diquark has a single channel
$\DQ_0 \to \bar{u}\; \bar{d}$, while the leptoquark has two final
states $\LQ_0 \to u\; e^- , \; d \nu_e$ with relative branching
fractions of 67\% and 33\%, respectively. For Case~B the channel
$D_0 \to D_{1/2}\; \wtd{\chi}_1^0$ opens up, but the decays to
fermion pairs continue to dominate. In the leptoquark case ${\rm
Br}(\LQ_0 \to \LQ_{1/2}\; \wtd{\chi}_1^0) = 15\%$ with the two SM
channels sharing the remainder in the ratio of 2:1. For the diquark
case ${\rm Br}(\DQ_0 \to \DQ_{1/2}\; \wtd{\chi}_1^0) = 5\%$ for
 $\lambda^9 = \lambda^{10} = 0.1$, with the
other 95\% accounted for by the two quark final state. For Case~C
all four neutralino final states are available, as is the decay
process $D_0 \to D_{1/2}\; \wtd{g}$. This latter mode is the
dominant channel, with a branching fraction of 87\% in the
leptoquark case and 77\% in the diquark case. The next largest
branching fraction for the leptoquark is ${\rm Br}(\LQ_0 \to
\LQ_{1/2}\; \wtd{\chi}_1^0) = 8\%$ and for the diquark is ${\rm
Br}(\DQ_0 \to \bar{u} \bar{d}) = 16\%$. The decay possibilities for
the cases~A-C are summarized in Table~\ref{decay_ferm}. As an
example we give in Figure~\ref{fig:Cbrats} the branching fraction
into the seven allowed final states for decays of the scalar
$\LQ_{0}$ as a function of the size of the leptoquark Yukawa
interaction. For convenience we here take $\lambda^6 = \lambda^7
\equiv \lambda_{\rm LQ}$. The vertical dotted line at $\lambda_{\rm
LQ} = 0.1$ defines the model we will consider in Section~\ref{sig}.

\begin{table}[thb]
\begin{center}
\begin{tabular}{|c||c|c||c|c||c|c|}
\multicolumn{1}{c}{}
 & \multicolumn{2}{c}{Case~A}
 & \multicolumn{2}{c}{Case~B}
 & \multicolumn{2}{c}{Case~C}\\ \hline
Decay & $D_{1/2}$ & $D_0^1$ & $D_{1/2}$ & $D_0^1$ & $D_{1/2}$ &
$D_0^1$ \\ \hline \hline
partner + $\wtd{\chi}^0_1$ &  &  &  & $\checkmark$ &  & $\checkmark$
\\ \hline
partner + $\wtd{\chi}^0_2$ &  &  &  &  &  & $\checkmark$
\\ \hline
partner + $\wtd{\chi}^0_3$ &  &  &  &  &  & $\checkmark$
\\ \hline
partner + $\wtd{\chi}^0_4$ &  &  &  &  &  & $\checkmark$
\\ \hline
partner + $\wtd{g}$ &  &  &  &  &  & $\checkmark\checkmark$
\\ \hline \hline
$\tilde{f} + f'$ & $\checkmark\checkmark_{\rm LQ}$ & NA &
$\checkmark\checkmark_{\rm LQ}$ & NA & $\checkmark\checkmark_{\rm
LQ}$ & NA \\
$f + f'$ & NA & $\checkmark\checkmark$ & NA & $\checkmark\checkmark$
& NA & $\checkmark$ \\ \hline \hline
$\wtd{\chi}^0_1 + f + f'$ & $\checkmark\checkmark_{\rm DQ}$ & &
$\checkmark\checkmark_{\rm DQ}$ & & $\checkmark\checkmark_{\rm DQ}$
& \\ \hline
%
%
\end{tabular}
\end{center}
{\caption{\label{decay_ferm}\footnotesize {\bf Decay possibilities
for cases~A-C in Table~\ref{benchmarks}}. A checkmark indicates an
allowed decay channel. The double checkmark indicates the channel
with the largest branching fraction. Note that this channel need not
be the same for the diquark and leptoquark scenarios. }}
\end{table}

\begin{figure}[t]
\begin{center}
\includegraphics[scale=0.48,angle=0]{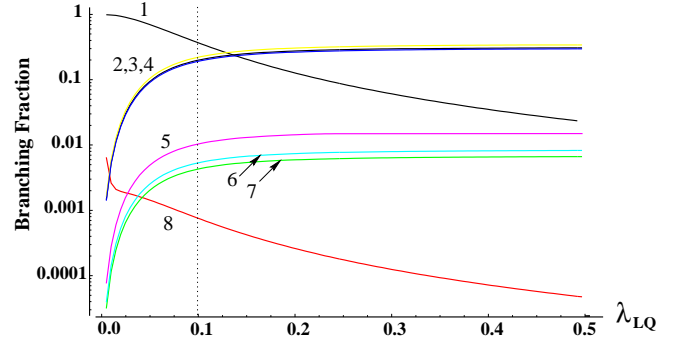}
\caption{\footnotesize \textbf{Branching ratios into various final
states for $\LQ_{1/2}$ decays in Scenario~D.} The branching fraction
into the eight allowed final states for the fermionic leptoquark are
given as a function of a (universal) Yukawa coupling $\lambda^6 =
\lambda^7 \equiv \lambda_{\rm LQ}$. These allowed final states are
as follows: (1) $D_{0}\,\tilde{\chi}_1^0$ (2,3,4)
$d\,\tilde{\nu}_e$, $u\,\tilde{e}_R$, $u\,\tilde{e}_L$ (5)
$e\,\tilde{u}_R$ (6) $e\,\tilde{u}_L$ (7) $\nu\,\tilde{d}_L$ (8)
$D_{0}\,\tilde{\chi}_2^0$.} \label{fig:Dbrats}
\end{center}
\end{figure}

The scalar partner is the LEP for cases~D and~E. Here the scalar
always decays via the Yukawa couplings of~(\ref{WLQ})
and~(\ref{WDQ}) to pairs of SM fermions. Case~D was designed to
allow the decays $D_{1/2} \to D_0\; \wtd{\chi}_{1,2}^0$, as well as
decays via the Yukawa interaction to $f\; \wtd{f}'$ final states.
The branching fraction for decay into the exotic scalar + LSP is
32\% for the leptoquark and 72\% for the diquark. In the latter
case, decays into $q \wtd{q}'$ final states are suppressed by the
relatively large squark mass scale. Final states such as $\wtd{d}\;
\nu_e$ and $\wtd{u}\; e^-$ are similarly suppressed for the
leptoquark. In this latter case the final states $d \;
\wtd{\nu}_e^L$, $u \; \wtd{e}_R^-$ and $u \; \wtd{e}_L^-$ each have
branching fractions in the 20-24\% range, making them competitive
with the $D_0\; \wtd{\chi}_{1}^0$ final state. The branching
fraction for $\LQ_{1/2}$ into the eight allowed final states are
plotted in Figure~\ref{fig:Dbrats} as a function of Yukawa coupling
$\lambda_{\rm LQ}$. Again we indicate the choice $\lambda_{\rm LQ} =
0.1$ by the vertical dotted line. For Case~E the very large mass
difference between the fermion and scalar LEP allows all decay
processes for both leptoquark and diquark. Here the dominant rate is
to $D_0\; \wtd{g}$ with branching fractions of 97\% and 94\%,
respectively. The decay possibilities for cases~D and~E are
summarized in Table~\ref{decay_scal}.

\begin{table}[thb]
\begin{center}
\begin{tabular}{|c||c|c||c|c|}
\multicolumn{1}{c}{}
 & \multicolumn{2}{c}{Case~D}
 & \multicolumn{2}{c}{Case~E} \\ \hline
Decay & $D_{1/2}$ & $D_0^1$ & $D_{1/2}$ & $D_0^1$ \\ \hline \hline
partner + $\wtd{\chi}^0_1$ & $\checkmark\checkmark$ &  & $\checkmark$ & \\
\hline
partner + $\wtd{\chi}^0_2$ & $\checkmark$ &  & $\checkmark$ & \\
\hline
partner + $\wtd{\chi}^0_3$ &  &  & $\checkmark$ & \\
\hline
partner + $\wtd{\chi}^0_4$ &  &  & $\checkmark$ &  \\
\hline
partner + $\wtd{g}$ &  &  & $\checkmark\checkmark$ &  \\
\hline \hline
$\tilde{f} + f'$ & $\checkmark$ & NA & $\checkmark$ & NA \\
$f + f'$ & NA & $\checkmark\checkmark$
& NA & $\checkmark\checkmark$ \\
\hline
\end{tabular}
\end{center}
{\caption{\label{decay_scal}\footnotesize {\bf Decay possibilities
for cases~D and~E in Table~\ref{benchmarks}}. A checkmark indicates
an allowed decay channel. The double checkmark indicates the channel
with the largest branching fraction. }}
\end{table}

Each individual decay channel was given its own position in unused
entries in the {\tt BRAT} and {\tt KFDP} common blocks in {\tt
PYTHIA}. Two-body decays are computed at initialization by the new
routine {\tt PYEXDC} on the basis of user-designated Yukawa
interactions and particle masses. Three-body decays require a more
sophisticated treatment due to the numerical phase space
integrations required. For these specific decays the result of these
numerical integrations are given directly to {\tt PYTHIA} for the
five scenario points in Table~\ref{benchmarks}.

\subsection{Collider Signatures}
\label{sig}

We now turn our attention to the signature for these exotic
supermultiplets at the CERN LHC. We are interested in what can be
learned about the exotic component of the SUSY signal in the
earliest stages of LHC data collection and analysis. Hence we will
study \lumint~of simulated signal data, where the signal here
includes both the events associated with the SPS~1a benchmark point
and one of our five exotic scenarios.
Signatures involving the diquark tend to be multi-jet events with
slightly less missing transverse energy than a ``typical'' MSSM SUSY
event. As we will not be simulating Standard Model backgrounds for
QCD jet events, making meaningful statements about
signal-to-background estimations is difficult, even with a full
detector simulation. A realistic study of such exotics would require
a more sophisticated treatment of SM multi-jet backgrounds than we
are providing in the current work. Furthermore, an additional
technical difficulty arises in this case when hadronizing the
out-going partons in {\tt PYTHIA} when non-standard ({\em i.e.}
non-planar) color flows are involved.\footnote{Unfortunately, we
have yet to find an adequate solution to this problem, which (to the
authors' knowledge) afflicts all publicly-available event generation
software.} For these two reasons we will defer the treatment of the
case of diquark couplings for the exotic multiplets to a subsequent
publication, when the technical issues mentioned here can be more
adequately resolved.

For the leptoquark scenario the Standard Model backgrounds are
similar to those of typical SUSY events and have been studied
elsewhere~\cite{Baer:1995nq,Baer:1995va,Hinchliffe:1996iu}. In this
paper we will rely on those estimates to allow us to discuss the
collider signatures without computing the relevant background
samples. Nevertheless, a sample of 200,000 $t\,\bar{t}$ and 40,000
$W^+\,W^-$/$W^{\pm}\, Z$/$Z\, Z$ events were generated for examining
the efficiency of our cuts in isolating the exotic leptoquark signal
from both the Standard Model and MSSM backgrounds. We proceed by
simulating the events with {\tt PYTHIA} and then pass the events to
{\tt PGS}~\cite{PGS} to simulate the detector response. The detector
parameters are those of the hybrid ATLAS/CMS detector used by the
LHC Olympics~\cite{LHCO}. As we do not wish to concern ourselves
unduly with detailed issues of triggering, we have chosen to utilize
the simplified triggering algorithm from version 3.0 of the {\tt
PGS} package. More specifically, we will analyze events for which
the quantity $H_T^{\rm trig} \geq 150 \GeV$, where
\begin{equation} H_T^{\rm trig} \equiv 5 \times \sum_{\rm lep}
|p_T^{\rm lep}| + 0.2 \times \sum_{\rm tau} |p_T^{\rm tau}| + 0.2
\times \sum_{\rm jet} |p_T^{\rm jet}| + |\not{\hspace{-.05in}{E_T}}|
\, . \label{trigger} \end{equation}
The objects appearing in the summation above are {\em candidate}
objects; {\em e.g.} the electrons and muons in the first term
in~(\ref{trigger}) are trigger-level electrons and muons. The taus
in~(\ref{trigger}) are hadronically-decaying tau-candidates. Crudely
speaking, electron and muon candidates must have $p_T > 10 \GeV$,
hadronically-decaying tau candidates and jet candidates must have
$p_T > 100 \GeV$ and $\met$ must be greater than 50 GeV to appear in
the trigger function.

\subsubsection{Can the new physics be definitively detected?}

As mentioned in the previous subsection, the spectrum for Snowmass
Point~1a gives rise to signatures that have been extensively studied
in the past. The presence of excesses above Standard Model
predictions can be definitively established in multiple channels in
this scenario for even a handful of inverse femptobarns of
integrated luminosity. The most important of these channels are
multi-jet final states, multi-jets plus one or more leptons, events
with same-sign pairs of leptons and any number of jets, trilepton
events, and events with three taus in the final
state~\cite{Barnett:1987dj,Baer:1991xs,Kane:2002qp}. All of these
cases are for events accompanied by large $\met$ values.

\begin{table}[t] \begin{center} \begin{tabular}{|c|c|c|}
\parbox{1.0in}{Sample} & \parbox{1.0in}{Accepted Events} & \parbox{1.0in}{Percentage}
\\ \hline
SPS~1a & 185,544 & 93.1\% \\ \hline
Case~A & 161,284 & 85.0\% \\ Case~B & 156,020 & 84.7\% \\
Case~C & 152,342 & 84.5\% \\ Case~D & 11,589 & 96.8\% \\
Case~E & 17.921 & 94.0\% \\ \hline
\end{tabular}
\end{center}
{\caption{\label{eventrate}\footnotesize {\bf Number of events in
\lumint~of integrated luminosity}. For our baseline model SPS~1a and
each of the five leptoquark cases \lumint~of data was simulated at
the LHC. The second column gives the number of these simulated
events for which the $H_T^{\rm trig}$ variable was over 150~GeV. The
third column gives the fraction of generated events which pass this
crude trigger threshold.}}
\end{table}

We expect the additional contribution from the iso\-singlet exotics to
add to these event rates -- making the discovery of ``new physics''
even more rapid. In \lumint~of integrated luminosity 185,544~MSSM
events from SPS~1a passed our low-level trigger function (an
acceptance rate of about 93\%). The corresponding number of accepted
events for the five exotic scenarios is given in
Table~\ref{eventrate}. The relatively large production rate for the
light exotic fermion in cases~A-C nearly doubles the new physics
sample, while the light-scalar scenarios provide only a small
additional contribution. Not all of these additional events,
however, will enter the SUSY search analyses.

\begin{figure}[t]
\begin{center}
\includegraphics[scale=0.46,angle=0]{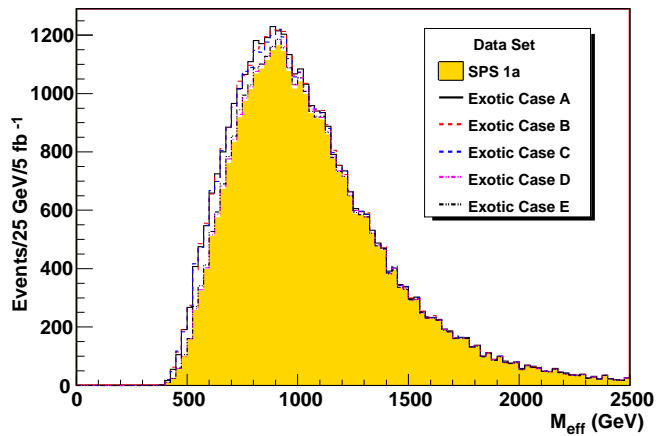}
\caption{\footnotesize \textbf{Effective mass distribution for
multi-jet (non-leptonic) events.} The effective mass
variable~(\ref{meff}) for the four hardest jets in events with no
leptons and $\met \geq 100 \GeV$. The five curves represent SPS~1a
together with each of the five leptoquark cases. The bulk of the
events passing the selection criteria are from SPS~1a, resulting in
almost no distinction between SPS~1a and the five exotic scenarios.}
\label{fig:meff}
\end{center}
\end{figure}

Let us consider the bluntest, but most effective, tool for
establishing the presence of beyond-the-Standard Model physics: the
$M_{\rm eff}$ analysis for events with multiple high-$p_T$ jets and
large $\met$~\cite{Baer:1995nq}. Following~\cite{Hinchliffe:1996iu}
we select events with at least four jets, no isolated leptons with
$p_T \geq 20 \GeV$, transverse sphericity $S > 0.2$, and $\met \geq
100 \GeV$. We require the $p_T$ of the hardest jet to be at least
100~GeV and the $p_T$ of the four hardest jets to each be above
50~GeV. The variable $M_{\rm eff}$ is given by the scalar sum
\begin{equation}
M_{\rm eff} = p_{T,1} + p_{T,2} + p_{T,3} + p_{T,4} + \met ,
\label{meff} \end{equation}
where $p_{T,i}$ is the transverse momentum of the $i$-th jet, ranked
in order of $p_T$ value. We plot this quantity for the SPS~1a model
together with each of the five exotic scenarios in
Figure~\ref{fig:meff}. For mass bins with $M_{\rm eff} \gappeq 600
\GeV$ a clear excess over the rapidly-falling Standard Model
background should be visible.

\begin{figure}[t]
\begin{center}
\includegraphics[scale=0.46,angle=0]{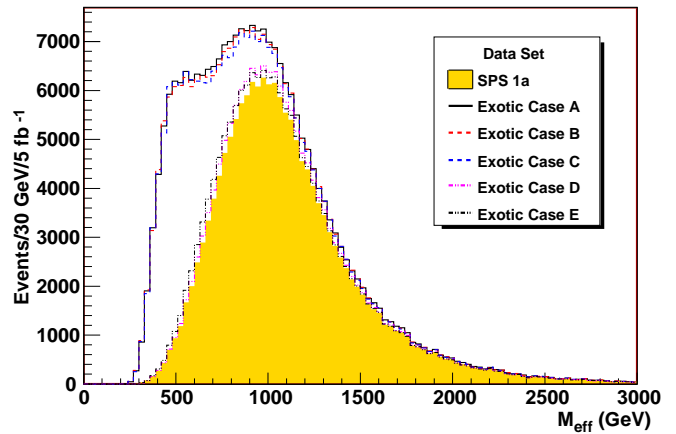}
\caption{\footnotesize \textbf{Effective mass distribution for all
high-$p_T$ objects in events with jets plus leptons.} The same
variable as Figure~\ref{fig:meff} but now formed from all high-$p_T$
objects with at least two jets and any number of leptons. In this
case the additional contribution from pair production of exotic
fermions is apparent in fermion-LEP cases A-C.} \label{fig:meff2}
\end{center}
\end{figure}

\begin{figure}[t]
\begin{center}
\includegraphics[scale=0.46,angle=0]{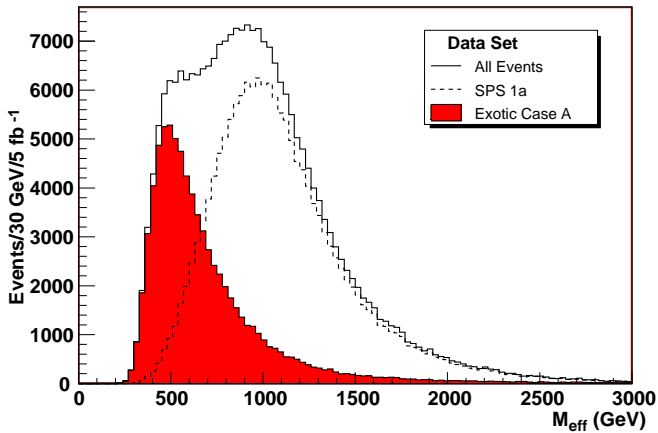}
\caption{\footnotesize \textbf{Breakdown of the effective mass
distribution for exotic case~A.} The contribution of exotic states
to the generalized $M_{\rm eff}$ variable is clearly separated from
the contributions of standard MSSM squarks and gluinos. A more
thorough analysis of Standard Model backgrounds is necessary,
however, to conclude that this structure in the low-mass bins is
experimentally observable.} \label{fig:meff3}
\end{center}
\end{figure}

Note, however, that the additional exotic component adds very little
to the event sample in this analysis. While 33,836 events in the
SPS~1a sample pass the above event selection cuts, less than 3,000
events pass them in cases~A-C, while for cases~D and~E only 410 and
490 events, respectively, enter the analysis. Even if we allow
events with any number of high-$p_T$ leptons we increase the SPS~1a
acceptance by a factor of 1.5, but for the exotic samples the
increase is only a modest factor of 3 for the fermion LEP cases and
2 for the scalar LEP cases. The most severe cuts turn out to be the
requirement of {\em four} hard jets and the cut on $\met$. We prefer
not to relax the $\met > 100 \GeV$ cut without a thorough
understanding of the Standard Model backgrounds. But in
Figure~\ref{fig:meff2} we plot the quantity $M_{\rm eff}$ for {\em
all objects} in events with at least {\em two} jets, each with
$p_T^{\rm jet} \geq 50 \GeV$ and any number of leptons (the other
cuts remaining the same as before). The scalar LEP cases continue to
contribute very little to the analysis, but now some structure is
evident in the $M_{\rm eff}$ distribution for the fermionic LEP
cases, albeit for low-energy bins where Standard Model processes
will likely be dominant. In Figure~\ref{fig:meff3} we focus on
Case~A where we show the individual contributions to the $M_{\rm
eff}$ variable of Figure~\ref{fig:meff2} from SPS~1a and from the
fermionic LEP Case~A. The two distributions are clearly separated,
with the SPS~1a peak representing squark and gluino production while
the peak for exotic Case~A represents fermion LEP pair production.
Whether or not the peak in this distribution at lower $M_{\rm eff}$
values is observable over the Standard Model background is an open
question that will require further study with full background
simulation.

\begin{figure}[t]
\begin{center}
\includegraphics[scale=0.5,angle=0]{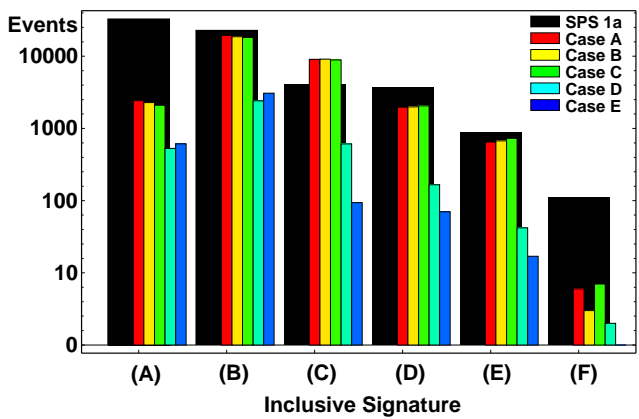}
\caption{\footnotesize \textbf{Number of events of different
inclusive signatures.} Six very inclusive signatures that may serve
as ``discovery modes'' for supersymmetric new physics are presented.
These signatures are: (A) multi-jet events, (B) jets + leptons, (C)
OS dileptons, (D) SS dileptons, (E) trileptons and (F) three-tau
events. All signatures have at least two high-$p_T$ jets and large
$\met$. Note the logarithmic scale of the vertical axis.}
\label{fig:inclusives}
\end{center}
\end{figure}

As our exotic signal will generally involve one or more hard leptons
in the final state we might expect that the inclusive signatures
involving leptons will be more highly augmented over the SPS~1a case
alone. In Figure~\ref{fig:inclusives} we give the event rate for the
six key inclusive signatures studied
in~\cite{Baer:1995nq,Baer:1995va,Kane:2002qp} for SPS~1a and the
five exotic scenarios of Table~\ref{benchmarks}. The first inclusive
signature involves the event sample for the $M_{\rm eff}$ analysis:
multi-jet events with large $\met$ and no isolated leptons. The
other five of these discovery modes involve high-$p_T$ leptons. They
are jet events with single isolated leptons, events with jets and
opposite-sign (OS) dileptons, events with jets and same-sign (SS)
dileptons, events with three leptons and jets and events with three
high-$p_T$ taus. All of these signatures involve large missing
transverse energy.

In contrast to Ref.~\onlinecite{Kane:2002qp} we give numbers of
events of these signature topologies after applying a set of cuts
designed to reduce the Standard Model background. These cuts are
based on those of Refs.~\onlinecite{Baer:1995nq,Baer:1995va}. All
events in Figure~\ref{fig:inclusives} have $\met \geq 100 \GeV$ and
$n_{\rm jet} \geq 2$, except for the multi-jet category where we
require $n_{\rm jet} \geq 3$. For the multi-jet case we require the
three hardest jets to have $p_T^{\rm jet} \geq 100 \GeV$. For the
other five signatures we require $p_T^{\rm jet} \geq 100 \GeV$ for
the hardest jet and $p_T^{\rm jet} \geq 50 \GeV$ for the
second-hardest jet. All events are required to have a transverse
sphericity of $S \geq 0.2$. For the multi-jet case we veto events
with leptons having $|\eta|\leq 2.5$ and $p_T^{\rm lep} \geq 20
\GeV$. For the topologies involving leptons we instead {\em require}
that the leptons (or taus) be isolated and have $p_T^{\rm lep} \geq
20 \GeV$.

As expected, the multi-jet channel is overwhelmingly populated with
events from the MSSM states. Event rates for the leptonic final
states for cases~A-C (with the fermion LEP) can be competitive with
those of standard SUSY -- particularly for the OS dilepton case
where the large production cross-section for fermionic LEPs couples
with the unit branching ratio for these states to quark + slepton.
For cases~D and~E (with the scalar LEP) the exotic contribution to
these inclusive signatures is a very small fraction of the total
event rate. Nevertheless, in a world with {\em only} the
supersymmetric exotic sector the scalar-LEP models are likely
discoverable on their own in \lumint~of data in the trilepton and
jets + lepton channels (particularly if judicious cuts on the size
of $\met$ are made in these events). The fermionic LEP models give
strong signals in all leptonic channels. When combined with the
large rate from SPS~1A it is clear that robust ``discovery'' modes
for new physics will be present across the board, even in the early
stages of LHC running.

\subsubsection{Can the exotic component be extracted?}

Observation of the combined event rates in the six channels of
Figure~\ref{fig:inclusives} should give clear evidence of new
physics at the LHC, and should furthermore strongly suggest
supersymmetry as the source of that new physics signal. Yet it is
not immediately clear that the exotic component -- above and beyond
the spectrum of the MSSM -- would be recognized merely from the
inclusive signature counts. Taking into account the logarithmic
scale in Figure~\ref{fig:inclusives} we note that the MSSM processes
still dominate most of the inclusive signatures, particularly for
exotic cases~D and~E. Nevertheless, the signal is clearly ``lepton
rich'' relative to SPS~1a alone. Presumably an attempt to fit
initial data to paradigms such as minimal supergravity would be
unable to reproduce the large rate for multi-lepton
events.\footnote{It would be an interesting exercise to attempt such
a fit within the context of a suitably general MSSM framework, such
as the ``minimal reasonable model'' of Brhlik and
Kane~\cite{Brhlik:1998gu}.} The first suggestion of beyond-the-MSSM
physics may therefore arise from fits to the data which suggest
extremely light slepton masses (say, below the bounds set by direct
search limits) or anomalously large branching fractions of gauginos
to leptonic final states.

One can find additional hints from the data itself. We have already
seen that plotting the effective mass of all objects in events with
at least 100 GeV of $\met$, at least two jets of $p_T^{\rm jet} \geq
50 \GeV$ and any number of leptons produces a structure indicating
some subset of events are {\em not} arising from gluino and/or
squark production. Yet these low effective-mass bins may not be
visible above the Standard Model background when the cuts of
Ref.~\onlinecite{Baer:1995nq} are loosened. Indeed, the rough
topology of events arising from pair production of leptoquark
$D_{1/2}$ pairs is not unlike top quark events; cuts designed to
reduce the rather large $t\,\bar{t}$ background at the LHC will
therefore tend to reduce the exotic signal in these channels as
well.

\begin{figure}[t]
\begin{center}
\includegraphics[scale=0.46,angle=0]{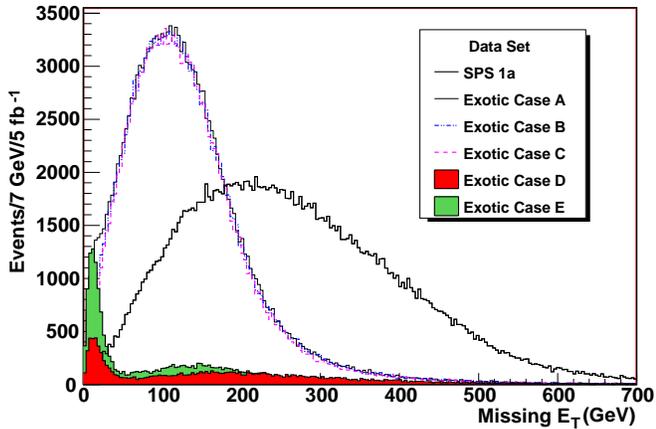}
\caption{\footnotesize \textbf{Missing transverse energy
distribution.} Total $\met$ in all events is plotted independently
for the SPS~1a data sample and the five exotic leptoquark cases. In
fermionic LEP models (open histograms) the exotic fermions decay
through cascade processes ending in the neutralino LSP. For scalar
LEP models (filled histograms) there is very little $\met$ for
scalar production, and very few events involving the much heavier
exotic fermion.} \label{fig:met}
\end{center}
\end{figure}

Nevertheless, some basic observations may suggest a leptoquark
interpretation for at least some of the signal events. Those events
arising from either fermionic or scalar leptoquarks distinguish
themselves from those arising from the MSSM states of SPS~1a in
three main ways: (1) lower jet multiplicities, (2) higher lepton
multiplicities and (3) lower values of $\met$. In
Figure~\ref{fig:met} we plot the distribution of missing transverse
energy in all events for each of our data samples. The standard MSSM
processes give rise to a broad distribution which peaks at
relatively large $\met$ values. For example, requiring a $\met$ cut
of 100 GeV only eliminates 9\% of the data sample (as opposed to
75\% of our $t\,\bar{t}$ sample and 87\% of our diboson sample). In
contrast, the fermionic LEP cases give a much sharper peak at
roughly half the value of the SPS~1a case. Nearly 50\% of these
events have $\met < 100 \GeV$. For the scalar LEP cases we have a
sizable production of the scalar exotic, which decays directly to
two Standard Model fermions with almost no missing energy. The long
shallow tails are from the production and decay of the much heavier
exotic fermion. Missing transverse energy in Standard Model events
tends to peak at or below 50 GeV and fall rapidly. This suggests
that at least for the fermionic LEP cases, the bimodal distribution
in $\met$ for the joint case of MSSM + exotics should indicate two
sets of fields being produced, both with production cross-sections
suggestive of strongly-interacting states, but with different mass
scales and decay topologies.

\begin{figure}[t]
\begin{center}
\includegraphics[scale=0.5,angle=0]{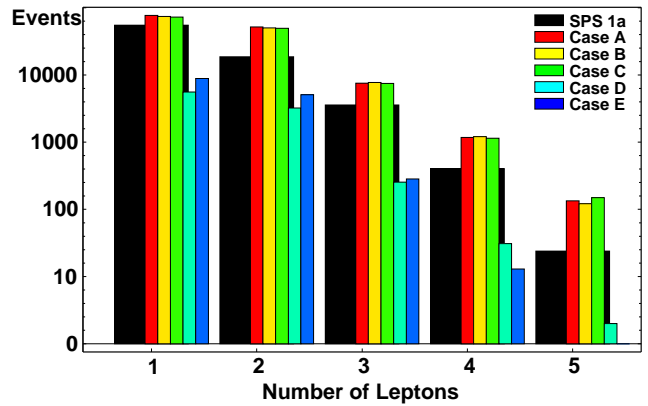}
\caption{\footnotesize \textbf{Lepton multiplicities.} Numbers of
events with one to five leptons, each having at least 20 GeV of
transverse momentum, for SPS~1a and the five exotic leptoquark
scenarios. Note the logarithmic vertical axis.} \label{fig:lepmult}
\end{center}
\end{figure}

\begin{figure}[t]
\begin{center}
\includegraphics[scale=0.46,angle=0]{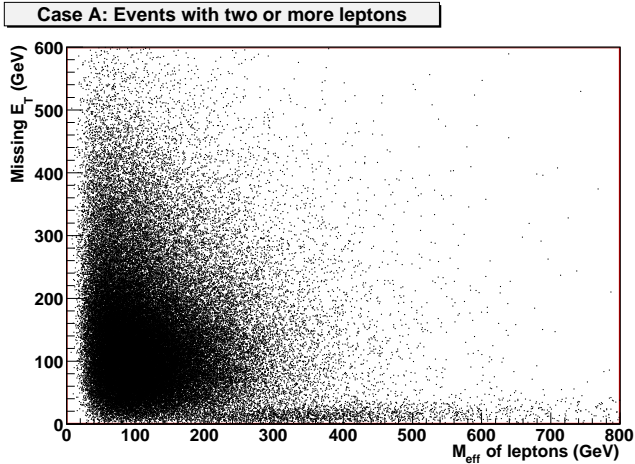}
\caption{\footnotesize \textbf{Scalar sum of lepton $p_T$ values
versus $\met$ for events in SPS~1a + exotic case~A.} Events in this
data set were required to have two isolated leptons but no other
cuts were made. The events at low values of $\met$ and large lepton
$p_T$ values are events from exotic leptoquark production,
particularly scalar pair production. } \label{fig:scatter}
\end{center}
\end{figure}

In Figure~\ref{fig:lepmult} we give the typical number of leptons
per event for the individual data samples. In each case the leptons
are required to have at least 20 GeV of transverse momentum, but no
other cuts are applied to the events. The exotic leptoquark samples
clearly favor higher lepton multiplicities, but the characteristics
of these leptons themselves (such as $M_{\rm eff}$ computed solely
from leading leptons, or the $p_T$ distribution of the hardest
lepton) do not differentiate between the SPS~1a events and the
exotic events. The experimentalist investigating these events will
find no peaks in the invariant mass distributions of various pairs
of leptons. However, the events with multiple high-$p_T$ leptons
tend to be those events with less $\met$. Consider, for example,
exotic case~A in conjunction with the SPS~1a model. In
Figure~\ref{fig:scatter} the scalar sum of $p_{T}^{\rm lep}$ values
for all leptons in events with at least two isolated leptons is
plotted versus the missing transverse energy in those events. It is
clear from Figure~\ref{fig:scatter} that there is some rough
anti-correlation between the energy accounted for by leptons and the
amount of energy carried away by LSPs. The feature visible along the
horizontal axis in the figure (that is, for extremely small values
of $\met$) with extremely energetic leptons are events involving the
pair production of scalar leptoquarks -- which have a sizable event
rate for the 367~GeV scalar mass of this benchmark point. The events
in this part of the scatter plot also appear in the low~$\met$ bins
in Figure~\ref{fig:met} for case~A. All of the events in
Figure~\ref{fig:scatter} have high-$p_T$ jets, yet no such
anti-correlation or low~$\met$ feature is apparent in the analogous
scatter plot involving jet $p_T$ values.

At this stage, if we were confronted with cases A-C we might
conclude that in addition to squarks and gluinos (with a mass-scale
of approximately 600-800 GeV) we are also producing a second class
of states with a mass-scale of approximately 300-500 GeV. The
overall rate for pair production of new states of this mass is
consistent with those states being fermions which interact through
the strong interaction. Events of this class occur with smaller
amounts of $\met$ than the squark/qluino events, and are associated
with increased numbers of leptons and fewer jets. The association of
strongly-coupled physics with prompt leptons and smaller but still
significant $\met$ signals {\em may} suggest a supersymmetric
leptoquark interpretation. For cases~D and~E, where the event rates
are lower, the $\met$ signal is muted or non-existent and the
Standard Model backgrounds are more problematic; it is unlikely that
any deviation from the MSSM spectrum would be apparent given the
crude observations presented here.

We have argued that \lumint~of integrated luminosity is likely
sufficient to reveal the presence of supersymmetric new physics. In
the three cases involving a light exotic fermion it is also likely
sufficient to reveal one or more objects with a production rate
indicative of strong coupling but with decay chains in which
high-$p_T$ leptons are prominent. Even in cases~D and~E for which
there will be little evidence for an exotic component to the SUSY
signal, it is still likely that experimental searches for
leptoquarks will be performed nonetheless.

\begin{figure}[t]
\begin{center}
\includegraphics[scale=0.46,angle=0]{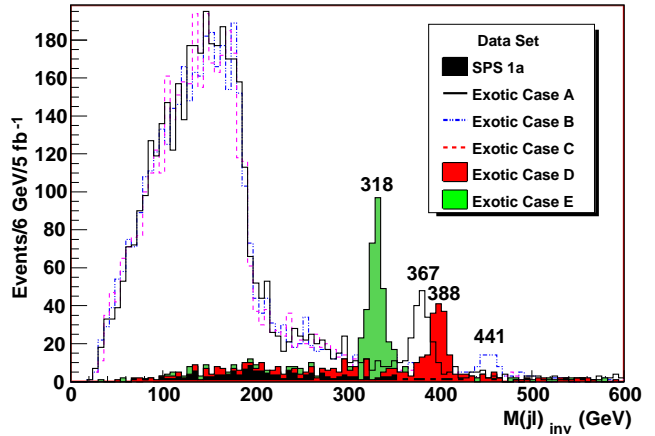}
\caption{\footnotesize \textbf{Invariant mass of hardest lepton
paired with softest jet in two jet, OS dilepton events.} Precisely
two jets, neither being B-tagged, were required, as were two
opposite-sign leptons. For cases A~C, with significant fermionic LEP
production, the end-point in the jet-lepton invariant mass
distribution at 200~GeV correctly gives the mass difference
$M_{D_{1/2}}- M_{\LSP}$. For the four cases where scalar production
was non-negligible a mass peak can be reconstructed near the
physical mass value for the lightest scalar.} \label{fig:jjll}
\end{center}
\end{figure}

In Figure~\ref{fig:jjll} we plot the invariant mass distribution of
the hardest lepton and softest jet in events with precisely two jets
and two (opposite-sign) leptons. Jets were required to have at least
50 GeV of transverse momentum and events were vetoed if either jet
was B-tagged. A cut was made on the $p_T$ of the leading lepton of
50~GeV, and 20~GeV for the trailing lepton. Finally, we require the
events to be somewhat collimated along the event axis, so we require
the transverse sphericity to be no greater than~0.7. This final cut
significantly reduced the contamination from both Standard Model
processes and SPS~1a events (an acceptance rate of approximately
0.04\% for each).

For cases~A-C the invariant mass of the jet/lepton pair shows an
end-point just below 200~GeV. This correctly measures the mass
combination
\begin{equation}
M_{\rm inv}^{\rm edge}(\ell \, j) = \sqrt{\frac{(M^2_{D_{1/2}} -
M_{\tilde{\ell}}^2)(M^2_{\tilde{\ell}} -
M_{\chi_1^0}^2)}{M^2_{\tilde{\ell}}}} \, , \label{edge}
\end{equation}
via the on-shell cascade decay $D_{1/2} \to q \tilde{\ell} \to q
\ell \chi_1^0$. For cases A-C this happens to be very near the mass
difference between the fermionic LEP and the lightest neutralino.
Mass peaks arising from the scalar pair production with $D_0 \to q
\ell$ can be reconstructed for all scenarios in which there is
significant scalar production (case~C had only 38 scalar events in
\lumint~of data). The true mass value for the lighter scalar is
given over the corresponding peak in Figure~\ref{fig:jjll}. We note
that if a cut on missing energy of $\met \geq 50 \GeV$ were applied,
the scalar mass peaks would vanish from the distributions in
Figure~\ref{fig:jjll}, though the end-point in the distribution
associated with fermion pair-production would still be visible.

These peaks can be isolated and sharpened by making stricter cuts on
the data set, such as demanding $\met \leq 25 \GeV$, requiring the
scalar sum of $p_T$ values from the two jets and two leptons sum to
at least 400~GeV, and requiring the invariant mass of the lepton
pair to be at least 100~GeV. An important cross-check is to find the
same peak in the jet/lepton invariant mass distribution in
associated production of scalar leptoquarks through the process
$g\,q \to D_0 q$. We can isolate this process by requiring (a) at
least two jets without B-tags, the hardest jet having at least
200~GeV of transverse momentum and all others having $p_T \geq 50
\GeV$, (b) precisely one isolated lepton with $p_T \geq 50 \GeV$,
and (c) $\met \leq 20 \GeV$. Pairing the second hardest jet with the
single lepton gives a clear peak at the same mass values as those in
Figure~\ref{fig:jjll}.

Thus, in every one of the scenarios of Table~\ref{benchmarks} there
should be at least one exotic state, and occasionally two such
states, which can be identified at the LHC -- even with limited
initial data. With additional statistics it should be possible to
measure the masses of low-lying scalar mass eigenstates in all five
scenarios. Reconstruction of cascade decays with additional
integrated luminosity should also allow a determination of the
exotic fermion mass in all five cases. We note, however, that the
presence of the (supersymmetric) exotic states can complicate the
analysis of the signal arising from MSSM states.

\begin{figure}[t]
\begin{center}
\includegraphics[scale=0.46,angle=0]{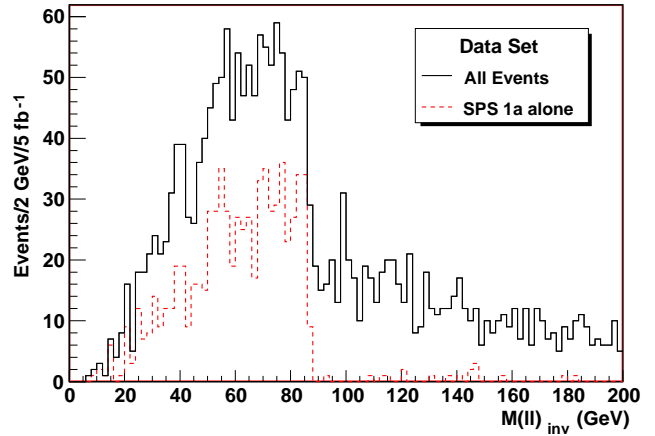}
\caption{\footnotesize \textbf{Flavor-subtracted dilepton invariant
mass distribution.} When flavor-universality is not present in the
leptoquark Yukawa couplings flavor-subtraction will not be adequate
to remove exotic fermion decays from $\chi_2^0$ decays.}
\label{fig:dilep}
\end{center}
\end{figure}

As the simplest example, we may consider the flavor-subtracted
dilepton invariant mass distribution for OS-dilepton events in
Snowmass point SPS~1a. For low-mass SUSY points such as this, even a
small data sample is sufficient to establish a sharp ``edge'' in the
weighted sum of distributions given by
\begin{equation} M_{\rm inv}(e^+ e^-) + M_{\rm inv}(\mu^+ \mu^-) - M_{\rm inv}(e^+
\mu^-)- M_{\rm inv}(\mu^+ e^-)\, . \label{flavsub} \end{equation}
The quantity in~(\ref{flavsub}) for SPS~1a is given by the dashed
line in Figure~\ref{fig:dilep}. The edge at $M_{\rm inv}(\ell^+
\ell^-) \simeq 90 \GeV$ is the result of the flavor-subtraction
procedure. When the pair of opposite-sign leptons arise from a
single decay chain of $\chi_2^0 \to \ell^{\pm} \tilde{\ell}^{\mp}
\to \ell^+ \ell^- \chi_1^0$, the two leptons are forced to be of the
same flavor. When the two leptons come from two independent decays (as
in the case of $t-\bar{t}$ production, or pair production of squarks
and gluinos) the probabilities for same-flavor and opposite-flavor
lepton pairs are roughly the same. The subtraction procedure
therefore efficiently isolates the exclusive decay chain for
$\chi_2^0$ decays and produces a sharp edge at the value
\begin{equation}
M_{\rm inv}^{\rm edge}(\ell^+ \ell^-) = \sqrt{\frac{(M^2_{\chi_2^0}
- M_{\tilde{\ell}}^2)(M^2_{\tilde{\ell}} -
M_{\chi_1^0}^2)}{M^2_{\tilde{\ell}}}} \, , \label{edge2}
\end{equation}
which happens to be very nearly the mass difference $M_{\chi_2^0} -
M_{\chi_1^0}$ in Snowmass point~1a.

Yet the effectiveness of this subtraction procedure is a result of
the flavor-democracy of Standard Model gauge interactions (which are
related to the decay rates of gauginos by supersymmetry). For our
exotic states these decays are governed by {\em Yukawa} interactions
which (up to model-dependent constraints from flavor-changing
processes discussed in Section~\ref{indirect}) are largely free
parameters. When the fermion LEP preferentially decays to one flavor
of lepton, for example, the subtraction technique will fail to
completely isolate gaugino decays. We show this effect by the solid
line in Figure~\ref{fig:dilep} where Snowmass point SPS~1a is
analyzed together with the exotic states of Case~A. In this case the
exotic is decaying exclusively to first-generation leptons. The
distribution still exhibits an edge (suggesting on-shell sleptons in
the decay chain), but the significance to this shape fitting will
diminish greatly without changes to the manner in which the gaugino
signal is extracted. This will necessarily degrade the accuracy with
which mass differences between low-lying superpartners and the LSP
can be determined.


\section{Conclusions} In this work we have provided a case study on how ``New Physics''
might appear in the early stages of the LHC experiment at CERN. If
this new physics is the minimal supersymmetric standard model (or
even more specifically, one of the specially-defined benchmark
points of this model that have been designed by theorists), then it
is very likely that the presence of the new physics and measurement
of many key properties will be straightforward. It is not at all
unlikely, however, that a supersymmetric world may be far from
minimal. Evidence in support of this comes both from top-down
constructions of GUT or string models as well as considerations of a
bottom-up nature which focus on naturalness of the electroweak
symmetry-breaking sector in
SUSY~\cite{Dine:2007xi,Brignole:2003cm,Casas:2003jx}.

We have presented here just such an alternative world that is very
much a plausible outcome of semi-realistic string constructions. The
common element of these models is the presence of $SU(3)$-charged
squarks and quarks which are isosinglets under weak $SU(2)$ and
carry equal and opposite hypercharge. Such states may present
themselves as new `partners' of Standard Model down-type quarks and
squarks, as leptoquarks or diquarks, or may have more exotic
properties which forbid renormalizable-level couplings altogether.
We have used the language of $E_6$ to characterize these states for
convenience, but $E_6$ GUT models are by no means the only way in
which such states can arise. The potential phenomenology of these
states is both rich and tightly constrained, providing an
interesting scenario for untangling high-scale physics from the
low-energy observables of the LHC.

If such states exist at LHC-accessible energies their presence will
most surely be detected. This was the conclusion of previous work
that considered such states in isolation, with only Standard Model
fields serving as `background.' In our work we have provided the
first analysis of such fields in the simultaneous presence of
`standard' superpartners, using the benchmark point of SPS~1a as our
example of simple SUSY models. The result of our study has been to
point out important consequences for experimenters hoping to observe
and measure the properties of new physics in {\em both} sectors. A
strong supersymmetric signal will not be obscured by these exotic
states, but recognition of the new states may not be immediate. In
the interim, we can expect fits to standard SUSY-breaking paradigms
such as minimal supergravity, to fit well initially. As the LHC
experiments integrate luminosity, we expect statistically
significant deviations from the predictions of the pure MSSM to
arise. In the leptoquark exotic case considered here that deviation
will first appear in events with both leptons and high-$p_T$ jets,
but with little missing transverse energy. We anticipate that astute
experimentalists will consider a leptoquark interpretation of these
`anomalous' events, but most SUSY analysis techniques (particularly
those which seek to isolate certain cascade decay chains and
therefore measure certain mass differences) will need significant
refinement. We hope that the work presented here will spur others to
provide a more sophisticated treatment of the quite-plausible
`what-if' scenario we provide, and to provide similar checks on the
robustness of long-established analysis techniques.

\begin{acknowledgments}
It is a pleasure to thank J.~Erler for useful discussions. 
This work was supported by
the U.S.~Department of Energy under Grant No.~DOE-EY-76-02-3071,
by National Science Foundation Grants PHY99-07949 and PHY-0503584, 
and by the Friends of the IAS.
\end{acknowledgments}

\myappendix

\mys{Alternative quasi-stable model} \label{qsalt} As mentioned in Section~\ref{E6case}, stability on the time-scale of
the detector at a hadron collider can be accomplished by forbidding
dimension-four operators that couple the exotic $SU(3)$
representations to the Standard Model. In Sections~\ref{E6case}
and~\ref{cosmo} this was illustrated by an $E_6$-motivated model in
which the dimension-four operators were forbidden by a combination
of $B$, $L$, and $U(1)_N$ symmetries. Here, we consider a toy model,
which was first constructed in \cite{Erler:2000wu},
in which this is achieved through a new $U(1)'$ gauge symmetry
alone. We will impose $U(1)'$ charge assignments such that our
exotics are stable at the renormalizable level but are allowed to
decay through higher-dimensional operators. To achieve this we
introduce an anomaly-free $U(1)'$ with chiral matter representations
which include two $\mathbf{5} + \overline{\mathbf{5}}$
representations of $SU(5)$ and various fields which will be singlets
under the Standard Model (SM) gauge group. Since $B$ and $L$ are not
imposed, we allow for the additional constraints imposed by the
non-observation of proton decay.

\begin{table}
{\begin{center}
\begin{tabular}{|c|c|c|c|c|c|} \hline
\parbox{1.6cm}{Field} & \parbox{1.0cm}{$Q_Y$} & \parbox{1.0cm}{$Q'$} &
\parbox{1.6cm}{Field} & \parbox{1.0cm}{$Q_Y$} & \parbox{1.0cm}{$Q'$}
\\ \hline
$3\times Q$ & 1/6 & 1 & $H_u$ & 1/2 & -2 \\ \hline
$3\times{u^c}$ & -2/3 & 1 & $H_d$ & -1/2 & -2 \\ \hline
$3\times{d^c}$ & 1/3 & 1 & $2\times D_t$ & -1/3 & -3 \\ \hline
$3\times L$ & -1/2 & 1 & $2\times{D}_t^c$ & 1/3 & -3 \\ \hline
$3\times{e^+}$ & 1 & 1 & $2\times L_s$& -1/2 & -2 \\ \hline
$3\times{\nu^c}$ & 0 & 1 & $2\times{L}_s^c$& 1/2 & -2 \\
\hline
$S$ & 0 & 4 & $T$ & 0 & 6 \\ \hline
\end{tabular}
\end{center}}
{\caption{\label{tbl:qsmodel} {\bf Particle Spectrum for the
alternative Quasi-Stable Exotic Model}. The fields of the
quasi-stable toy model are listed with hypercharge $U(1)_Y$ and
exotic $U(1)'$ charge assignments. The  new fields $L_s$ and $D_t$
come in two vector-like pairs. $D_t$ transforms as a charge $-1/3$
color triplet, while $L_s$  and $L_s^c$ can be interpreted as additional Higgs pairs
or as exotic lepton doublets. The new fields $S$ and $T$ are Standard Model
singlets.}}
\end{table}

The field content of the model is given in Table~\ref{tbl:qsmodel}.
The $U(1)'$ charge assignments are obtained by solving the anomaly
cancellation equations. For the $U(1)'$ charges listed below all the
mixed anomalies of $G_{\SM} \times U(1)'$ and $Y \times U(1)'^2$
vanish. Since we only include additional representations
$\mathbf{5}+\oline{\mathbf{5}}$, we will not modify the SM gauge
anomaly cancelations. However, additional SM singlets are still
needed to cancel the trace anomaly and the $U(1)'^3$ anomaly. There
is some freedom in choosing their $U(1)'$ charge assignments \cite{Erler:2000wu}. Note
that the model maintains the unification of the SM gauge couplings
at one loop.  In fact, complete unification can be achieved for the
appropriate rescaling of the $U(1)'$ gauge coupling $g'$. However,
there is no simple explanation of the $U(1)'$ charge quantization,
analogous to the $U(1)_Y$ charge quantization in ordinary grand
unification theories, since the $U(1)'$ charge assignments are just
numerical solutions to the anomaly cancelation equations.

Aside from the usual Yukawa interactions of the MSSM, the
superpotential terms allowed by the choice of $Q'$ charges in
Table~\ref{tbl:qsmodel} are given by
\begin{eqnarray}
{\rm dim4}&:& S H_u H_d,\; T D_t {D}_t^c,\; S L_s {L}_s^c,
\nonumber \\
%
& & H_d Q {d^c}, \; H_u Q {u^c},\; H_d L e^+ , \;
H_u L {\nu^c}, \nonumber \\
& & + (H_u \rightarrow L_s^c)\; + (H_d \rightarrow {L}_s),
\label{dim4} \\
{\rm dim5}&:& {D}_t^c Q H_d S, \; {D}_t^c Q L_s S, \;
{D}_t^c Q Q {u^c}, \nonumber \\
& & D_t {d^c} {\nu^c}{\nu^c}, \; D_t {u^c} e^+{\nu^c},\;
{D}_t^c Q L {\nu^c}, \nonumber \\
& & {D}_t^c{u^c}{u^c} e^+, \; D_t Q Q {\nu^c},\; {D}_t^c
{d^c}{u^c}{\nu^c} . \label{dim5}
\end{eqnarray}
The exotic particles $D_t$ and ${D}_t^c$ are stable at
renormalizable level: there is only one operator $T D_t  D_t^c$
involving $D_t$ that is allowed by the symmetry, and one can
therefore assign a conserved $D_t$ number. This operator can provide
a mass term for the exotic quarks should a $U(1)'$-breaking vacuum
expectation value $\lang T \rang$ develop. 
The $L_s$ and $L_s^c$ doublets can decay by the dimension 4 operators.

There are a number of $D_t$-number violating interactions allowed at
dimension~5, however. These operators also violate baryon and lepton
numbers, leading to proton decay. The operators can be divided into
two sets. The first two rows of~(\ref{dim5}) would allow $B$
conservation for the assignment $B=\frac{1}{3}(-\frac{1}{3})$ for
$D_t ({D}_t^c)$, while the last row of~(\ref{dim5}) would require
$B=-\frac{2}{3}(+\frac{2}{3})$. Clearly, if operators from both sets
are present $B$ will be violated and the proton will decay. However,
since the $D_t$-number is conserved at dimension-4 level (and the
$D_t$ is assumed to be heavier than the proton), the proton decay
rate $\Gamma_p$ is proportional to $M^{-4}_*$, where $1/M_*$ is the
scale multiplying the dimension-5 operators. Depending on which
operators are included, the dominant decay processes are $p \to
\pi^+ \nu $ ($\Delta_{B-L} = -2$) or $p \to \pi^+  \nu^c $
($\Delta_{B-L} = 0$), or the corresponding neutron decays $n \to
\pi^0 \nu (\nu^c)$.\footnote{Decays into $e^+$ are highly suppressed
because the operator ${D}_t^c {u^c}{u^c} e^+$ must be antisymmetric
in the ${u^c}{u^c}$ flavor indices, leading to decays including a
virtual $c$ or ${c^c}$ quark.} The experimental limits for these
processes are $\tau \ge 10^{31} {\rm yr}\ (p \to \pi^+ \nu)$ and
$\tau \ge 10^{32} {\rm yr}\ (n \to \pi^0
\nu)$~\cite{Eidelman:2004wy}. If the operator ${D}_t^c Q H_d S$ is
present, with the scalar components of $H_d$ and $S$ acquiring VEVs,
as well as $D_t Q Q {\nu^c}$ or ${D}_t^c {d^c}{u^c} {\nu^c}$, the
proton decay constraint is satisfied provided that $M \ge
10^{15}-10^{16} \GeV$. Alternatively, if ${D}_t^c Q H_d S$ is
replaced by ${D}_t^c Q Q {u^c}$, then $M \ge 10^{12}-10^{13}\GeV$ is
sufficient.

All of the dimension-five operators in~(\ref{dim5}) allow the $D_t$
to decay. However, up to counting factors the results are expected
to be similar to those for the $E_6$-motivated model discussed in
Sections~\ref{E6case} and~\ref{cosmo}.

\mys{Explicit Cross-Section and Partial Width Expressions}
\label{xsec}
\label{appendixB}
In keeping with the conventions of {\tt PYTHIA} we quote
parton-level cross sections for $2 \to 2$ processes as differential
cross-sections of the form $\diff \wh{\sigma}/\diff \hat{t}$. In
what follows the kinematic factor $\beta_{34}$ is given by
\begin{eqnarray}
\beta_{34} &=& \sqrt{\(1-\frac{m_3^2}{\hat{s}} -
\frac{m_4^2}{\hat{s}}\)^2 - \frac{4m_3^2 m_4^2}{\hat{s}^2}}
\nonumber \\ &\to& \sqrt{1-\frac{4M_D^2}{\hat{s}}} ,
\label{beta34} \end{eqnarray}
where the last relation holds when $m_3 = m_4 = M_D$. We will also
have occasion to use the quantity
\begin{equation}
{M}_{34}^2 = \frac{1}{2}(m_3^2 + m_4^2) -
\frac{1}{4\hat{s}}(m_3^2 - m_4^2)^2 \to M_D^2 .
\label{mass34} \end{equation}
In {\tt PYTHIA} the parton-level Mandelstam variables are given by
\begin{equation}
\hat{t},\hat{u} = -\frac{1}{2}\[ (\hat{s} - m_3^2 - m_4^2) \mp
\hat{s}\beta_{34}\cos\wh{\theta}\]
\end{equation}
where $\wh{\theta}$ is the polar angle of parton~3 in the c.m. frame
of the hard scattering.

\begin{widetext}
\subsection{Production Cross Sections}
\begin{itemize}
\item $q + \bar{q} \to D_{1/2} \oline{D}_{1/2}$

There are two parton-level sub-processes of relevance: the
$q$-$\bar{q}$ annihilation (two quarks of the same flavor) through
gluon exchange in the direct channel and a diagram with t-channel
exchange of a squark or slepton through the Yukawa interaction. The
differential cross-section for the leptoquark case is given by
\begin{eqnarray}
\frac{\diff \wh{\sigma}}{\diff \hat{t}} &=&  \frac{4\pi
\alpha_s^2}{9\hat{s}^2}
\[\frac{(\wh{t}-M_{34}^2)^2 + (\wh{u}-M_{34}^2)^2 +
2M_{34}^2\wh{s}}{\wh{s}^2}\] - \frac{8 \pi \alpha_s
\alpha_y}{9\hat{s}^2} \[\frac{\hat{s}M_{34}^2
 + (\hat{t} - M_{34}^2)^2}{\hat{s}(\hat{t}-m_{\tilde{e}_L}^2)}
 + \frac{\hat{s}M_{34}^2
 + (\hat{t} - M_{34}^2)^2}{\hat{s}(\hat{t}-m_{\tilde{e}_R}^2)}\]
 \nonumber \\
 & & + \frac{\pi\alpha_y^2}{\hat{s}^2} \[
 \frac{(\hat{t} - M_{34}^2)^2}{(\hat{t}-m_{\tilde{e}_L}^2)^2} +
 \frac{(\hat{t} - M_{34}^2)^2}{(\hat{t}-m_{\tilde{e}_R}^2)^2}\] +
 \frac{2 \pi \alpha_y^2}{\hat{s}^2}
 \frac{(\hat{t} - M_{34}^2)^2}{(\hat{t}-m_{\tilde{e}_L}^2)
 (\hat{t}-m_{\tilde{e}_R}^2)} \; ,
\label{ISUB325} \end{eqnarray}
where $\alpha_y = \lambda^2/4\pi$. For the diquark case we have
\begin{eqnarray}
\frac{\diff \wh{\sigma}}{\diff \hat{t}} &=&  \frac{4\pi
\alpha_s^2}{9\hat{s}^2}
\[\frac{(\wh{t}-M_{34}^2)^2 + (\wh{u}-M_{34}^2)^2 +
2M_{34}^2\wh{s}}{\wh{s}^2}\] + \frac{8 \pi \alpha_s
\alpha_y}{9\hat{s}^2} \[\frac{\hat{s}M_{34}^2
 + (\hat{t} - M_{34}^2)^2}{\hat{s}(\hat{t}-m_{\tilde{q}_L}^2)}
 + \frac{\hat{s}M_{34}^2
 + (\hat{t} - M_{34}^2)^2}{\hat{s}(\hat{t}-m_{\tilde{q}_R}^2)}\]
 \nonumber \\
 & & + \frac{4\pi\alpha_y^2}{3\hat{s}^2} \[
 \frac{(\hat{t} - M_{34}^2)^2}{(\hat{t}-m_{\tilde{q}_L}^2)^2} +
 \frac{(\hat{t} - M_{34}^2)^2}{(\hat{t}-m_{\tilde{q}_R}^2)^2}\] +
 \frac{8 \pi \alpha_y^2}{3\hat{s}^2}
 \frac{(\hat{t} - M_{34}^2)^2}{(\hat{t}-m_{\tilde{q}_L}^2)
 (\hat{t}-m_{\tilde{q}_R}^2)} .
\label{ISUB330} \end{eqnarray}
%
Note the following: (1) In doing the PYTHIA analysis we always take
$\lambda^6 = \lambda^7$, $\lambda^8=0$ and $\lambda^9 =
\lambda^{10}$. So expressions involving $\alpha_y$ should be
understood as involving the common value of this Yukawa interaction.
(2) For the Snowmass point we use as our superpartner spectrum, the
up and down type left-handed squarks have the same masses to within
a percent and up and down right-handed squarks have the same masses
to within less than a percent. Thus, in evaluating~(\ref{ISUB330})
PYTHIA takes just the (common) $\tilde{q}_L$ soft mass and
$\tilde{q}_R$ soft masses.

\item $g + g \to D_{1/2} \oline{D}_{1/2}$

There are two sub-diagrams for this process: the diagram with the
initial gluons leading to a gluon intermediate state and then to two
fermions, and the t-channel/u-channel exchange of the heavy fermion
$D_{1/2}$.
%
%
The differential cross-section is given by
\begin{eqnarray}
\frac{\diff \wh{\sigma}}{\diff \hat{t}} &=&
\frac{\pi\alpha_s^2}{6\hat{s}^2}\lbr
\frac{\hat{u}-M_{34}^2}{\hat{t}-M_{34}^2}
-\frac{9}{4}\frac{(\hat{u}-M_{34}^2)^2}{\hat{s}^2} +
\frac{9}{2}\frac{M_{34}^2}{\hat{s}}
\frac{(\hat{t}-M_{34}^2)(\hat{u}-M_{34}^2)-
M_{34}^2\hat{s}}{(\hat{t}-M_{34}^2)^2} \right. \nonumber \\
 & & \quad \quad \left. +\frac{M_{34}^2\hat{t}}{2(\hat{t}-M_{34}^2)^2}
-\frac{M_{34}^4}{\hat{s}(\hat{t}-M_{34}^2)} \rbr ,
\label{ISUB326} \end{eqnarray}
plus the equivalent expression with $\hat{t} \leftrightarrow
\hat{u}$.

\item $q + \bar{q} \to D_0 \oline{D}_0$

For this process there are two types of diagrams: $q$-$\bar{q}$
annihilation (two quarks of the same flavor) through gluon exchange
in the direct channel and t-channel exchange of a Standard Model
fermion.
%
%
As we allow our exotics to interact solely with first generation
Standard Model fermions, the appropriate differential cross-section
is
\begin{equation}
\frac{\diff \wh{\sigma}_0}{\diff \hat{t}} = \frac{\pi
\alpha_s^2}{9\hat{s}^2}\lbr \frac{\hat{s}(\hat{s}-4M_{34}^2) -
(\hat{u} - \hat{t})^2}{\hat{s}^2} \rbr
\label{ISUB327a} \end{equation}
when $q$,$\bar{q}$ are a flavor other than the first generation and
\begin{equation}
\frac{\diff \wh{\sigma}}{\diff \hat{t}} = \frac{\diff
\wh{\sigma}_0}{\diff \hat{t}} +
\frac{\pi\alpha_s\alpha_y}{18\hat{s}^2}\lbr \frac{(M_{34}^2 -
\hat{t})(\hat{u}-\hat{t})+\hat{s}(M_{34}^2 +
\hat{t})}{\hat{s}\hat{t}} \rbr \nonumber +
\frac{\pi\alpha_y^2}{8\hat{s}^2}\lbr \frac{-\hat{s}\hat{t}-(M_{34}^2
- \hat{t})^2}{\hat{t}^2} \rbr
\label{ISUB327b} \end{equation}
for first-generation $q$,$\bar{q}$.
%

\item $g + g \to D_0 + \oline{D}_0$

The QCD diagrams for the diquarks and leptoquarks are the same.
There are several sub-diagrams:
the diagram with the initial gluons leading to a gluon intermediate
state and then to two scalars, the t-channel/u-channel exchange of
the heavy scalar $D_0$, and the ``sea-gull'' four-point diagram. The
appropriate differential cross-section is given by
%
\begin{equation}
\frac{\diff \wh{\sigma}}{\diff \hat{t}} =  \frac{\pi
\alpha_s^2}{2\hat{s}^2}\lbr \frac{7}{48} +
\frac{3(\hat{u}-\hat{t})^2}{16\hat{s}^2} \rbr\lbr 1 + \frac{2M_D^2
\hat{t}}{(\hat{t}-M_D^2)^2} +
\frac{2M_D^2\hat{u}}{(\hat{u}-M_D^2)^2} +
\frac{4M_D^4}{(\hat{t}-M_D^2)(\hat{u}-M_D^2)} \rbr .
\label{ISUB328} \end{equation}
%

\item $q + g \to D_0 + f$

Single scalar leptoquarks and diquarks can be produced in
association with a lepton and quark, respectively. The diagram
involves $g q$ initial states, either a quark in the direct channel
or an exotic squark in the indirect (u) channel, and then a scalar and
fermion in the final state. One of the two vertices is then given by
QCD while the other is from the appropriate Yukawa coupling.
The expression for the leptoquark is given by
\begin{equation}
\frac{\diff \wh{\sigma}}{\diff \hat{t}} =
\frac{\pi\alpha_s\alpha_y}{6\hat{s}^2} \(-\frac{\hat{t}}{\hat{s}}\)
\frac{\hat{u}^2 + M_D^4}{(\hat{u}-M_D^2)^2} \; \; .
\label{ISUB329} \end{equation}
The equivalent expression for the diquark gets an extra factor of
two from the different color contraction.

\item $\bar{q} + \bar{q} \to \DQ_0$

This case is slightly different. What's coded into {\tt PYTHIA} is
$(\hat{s}/\pi)\diff \sigma$, and this is given in turn by
\begin{equation}
\frac{\hat{s}}{\pi} \diff \wh{\sigma} =
4\hat{s}\frac{4}{3}\frac{\lambda^2}{16\pi}
\frac{\sqrt{\hat{s}}\Gamma_{D_0}}{(\hat{s} - M_{D_0}^2)^2 +
\hat{s}\Gamma^2_{D_0}} .
\end{equation}
\end{itemize}

\subsection{Two-Body Decays}
\noindent For two-body decays of the form $a \to b+c$ let us define
the final momentum in the rest frame of $a$ as
\begin{equation}
p_f=\frac{[m_a^2-(m_b+m_c)^2]^{1/2}[m_a^2-(m_b-m_c)^2]^{1/2}}{2m_a}
. \label{pfinal} \end{equation}
Furthermore, we define the exotic scalar mixing matrix by the
relations
\begin{equation} D_0 = D_0^1 \cos \theta_D - D_0^2 \sin \theta_D \;
; \quad \quad D_0^{c*} = D_0^1 \sin \theta_D +D_0^2 \cos \theta_D .
\label{Dmix}
\end{equation}
Then the partial widths for the exotic decay channels considered are
given by the following:\footnote{We always give the spin-averaged
decay rate for the exotic fermions. For QCD production, this ignores
spin correlations between the produced $D$ and $D^c$. It also
neglects possible polarizations for leptoquark or diquark production
processes.}
\begin{itemize}
\item $D_{1/2}\to D_0^1+\tilde{g}$ 
\begin{equation} \Gamma = \frac{4}{3} \frac{\alpha_s}{2} p_f \left[ 1 +
\frac{m^2_{\tilde{g}}}{M_{D_{1/2}}^2} -
\frac{M_{D_0}^2}{M_{D_{1/2}}^2}  - 4 \cos \theta_D \sin \theta_D
\frac{m_{\tilde{g}}}{M_{D_{1/2}}} \right]
\end{equation}
where the $\frac{4}{3}$ is from color.

\item $D_{1/2}\to D_0^1+\chi_i^0$
\begin{equation} \Gamma = \frac{\alpha_2}{2}
|Q_D \tan\theta_W N_{i1}|^2 p_f \left[ 1 +
\frac{m^2_{\chi_i^0}}{M_{D_{1/2}}^2} -
\frac{M_{D_0}^2}{M_{D_{1/2}}^2} -4 \cos \theta_D \sin \theta_D
\frac{m_{\chi_i^0}}{M_{D_{1/2}}} \right]
\end{equation}
where $Q_D=-\frac{1}{3}$, $\alpha_2= \frac{g_2^2}{4\pi}$, and $N_{i1} (N_{i2})$ is the mixing
element relating $\chi_i^0$ to the bino (wino).
We  ignore contributions to decays from neutralinos
associated with the Standard Model singlet  $S$ (singlinos) or additional $Z'$
($Z'$-gauginos).
\item $D_0^1\to D_{1/2}+\tilde{g}$ 
\begin{equation} \Gamma = \frac{4}{3} \alpha_s p_f \left[ 1 -
\frac{m^2_{\tilde{g}}}{M_{D_{0}}^2} -
\frac{M_{D_{1/2}}^2}{M_{D_{0}}^2} + 4 \cos \theta_D \sin \theta_D
\frac{m_{\tilde{g}}M_{D_{1/2}}}{M^2_{D_{0}}} \right] .
\end{equation}
\item $D_0^1\to D_{1/2}+\chi_i^0$
\begin{equation} \Gamma = \alpha_2
|Q_D \tan\theta_W N_{i1}|^2 p_f \left[ 1 -
\frac{m^2_{\chi_i^0}}{M_{D_{0}}^2} -
\frac{M_{D_{1/2}}^2}{M_{D_{0}}^2} +4 \cos \theta_D \sin \theta_D
\frac{m_{\chi_i^0}M_{D_{1/2}}}{M^2_{D_{0}}} \right] .
\end{equation}
\end{itemize}
\end{widetext}

\begin{itemize}
\item For the process $D_0^1 \to f f'$, neglecting fermion masses and
taking $\lambda^6 = \lambda^7 = \lambda_{\sc LQ}$ and $\lambda^9 =
\lambda^{10} = \lambda_{\sc DQ}$, we have
\begin{equation}
\Gamma_i = \frac{d_i}{16\pi}M_{D_0} \end{equation}
with
\begin{equation} \left. \begin{array}{c} \LQ_0 \to u e^- \\
\LQ_0 \to d \nu_e \\ \DQ_0 \to \bar{u} \bar{d} \end{array} \rbr
\to \lbr \begin{array}{c} d_i = |\lambda_{\sc LQ}|^2 \\
d_i = \sin^2\theta_D|\lambda_{\sc LQ}|^2 \\ d_i =
2(1+3\cos^2\theta_D)|\lambda_{\sc DQ}|^2
\end{array} \right. . \end{equation}
\item
When kinematically allowed, the rate for the process $D_{1/2} \to f
\tilde{f}'$ is given in the same approximation above by
\begin{equation}
\Gamma_i = \frac{d_i}{16\pi} p_f \left[ 1 -
\frac{m^2_{\tilde{f}'}}{M^2_{D_{1/2}}}\right]
\end{equation}
with $d_i = |\lambda_{\sc LQ}|^2$ for all $\LQ_{1/2}$ decays and
$d_i = 8|\lambda_{\sc DQ}|^2$, $2|\lambda_{\sc DQ}|^2$ for
$\DQ_{1/2}$ decays to left-handed and right-handed squarks,
respectively.

\end{itemize}

\subsection{Three-Body Decays}
Finally, we give the rates for three-body decays. Let us keep only
the first generation of Standard Model fermions and ignore
$\tilde{f}_L \tilde{f}_R$ mixing, except for the exotic scalars
whose mixing is defined in (\ref{Dmix}).
We continue to ignore contributions to decays from singlinos or $Z'$-gauginos. We
will also ignore Higgsino contributions. We define the
quantities
\begin{equation}
\xDl^*=\xDr= \xdr = -\frac 1 3 \tan \theta_W N_{11} \end{equation}
\begin{equation} \xdl^* = - \frac 1 2 N_{12} + \frac 1 6  \tan \theta_W N_{11}
\end{equation}
\begin{equation} \xul^* = + \frac 1 2 N_{12} + \frac 1 6  \tan \theta_W N_{11}
\end{equation}
\begin{equation} \xur = + \frac 2 3  \tan \theta_W N_{11} \end{equation}
\begin{equation} \xel^* = - \frac 1 2 N_{12} - \frac 1 2  \tan \theta_W N_{11}
\end{equation}
\begin{equation} \xer = -  \tan \theta_W N_{11} \end{equation}
\begin{equation} \xnl^* = + \frac 1 2 N_{12} - \frac 1 2  \tan \theta_W N_{11} \, ,
\label{Xdef} \end{equation}
where $N_{11}$ and $N_{12}$ are the entries of the ($4\times4$)
neutralino mass matrix. In practice we will take $N_{11} = 1$ and
$N_{12}=0$.

Let us begin by considering the process $\DQ_{1/2} \to \chi_1^0 d^c
u^c$ via the operator with coefficient $\lambda^{10}$
in~(\ref{WDQ}). Diagrams leading to this final state can involve the
virtual $\DQ_0$ state or virtual right-handed squarks. The
amplitude-squared, summed over final colors and averaged over
initial spins, is given by
\begin{equation} |\oline{\mathcal{M}}|^2 = 4C |\lambda^{10}|^2 g_2^2
|\mathcal{A}|^2 \, , \label{msquared} \end{equation}
where $C=2$ is the color factor for this process and
\begin{widetext}
\begin{eqnarray}
|\mathcal{A}|^2 &=& |\ydr|^2 (p_{u^c} \cdot p_D)(p_{d^c} \cdot
p_{\chi}) + |\yur|^2 (p_{d^c} \cdot p_D)(p_{u^c} \cdot p_{\chi})
+\(|\yDl|^2+|\yDr|^2\) (p_{u^c} \cdot  p_{d^c}) ( p_{\chi} \cdot
 p_D) \nonumber \\
 & & -{\rm Re}(\ydr^*\yur)[(p_{u^c} \cdot p_D)(p_{d^c} \cdot p_{\chi})
 + (p_{u^c} \cdot p_{\chi})(p_{d^c} \cdot p_D)
 - (p_{u^c} \cdot p_{d^c})(p_{\chi} \cdot p_D)] \nonumber \\
 & & -{\rm Re}(\yDr^*\ydr)[(p_{u^c} \cdot p_D)(p_{d^c} \cdot p_{\chi})
 + (p_{u^c} \cdot p_{d^c})(p_{\chi} \cdot p_D)
 - (p_{u^c} \cdot p_{\chi})(p_{d^c} \cdot p_D)] \nonumber \\
 & & -{\rm Re}(\yDr^*\yur)[(p_{d^c} \cdot p_D)(p_{u^c} \cdot p_{\chi})
 + (p_{u^c} \cdot p_{d^c})(p_{\chi} \cdot p_D)
 - (p_{d^c} \cdot p_{\chi})(p_{u^c} \cdot p_D)] \nonumber \\
 & & + \left\{ 2 {\rm Re}(\yDl^*\yDr) -{\rm Re}(\yDl^*\ydr) - {\rm Re}(\yDl^*\yur)\right\}
 M_D M_\chi (p_{u^c} \cdot p_{d^c}) \, .
\label{amplitude} \end{eqnarray}
The $Y_i$ are given by the expressions
\begin{equation}
\ydr = \frac{\xdr}{(p_{u^c}-p_D)^2 - m^2_{\tilde{d}_R}}\, , \qquad
\yur = \frac{\xur}{(p_{d^c}-p_D)^2 - m^2_{\tilde{u}_R}} \label{Yq}
\end{equation}
\begin{eqnarray}
\yDr &=& \xDr \left[ \frac{\sin^2\theta_D}{(p_{\chi}-p_D)^2 -
m^2_{D_0^1}}+ \frac{\cos^2\theta_D}{(p_{\chi}-p_D)^2 -
m^2_{D_0^2}}\right] \nonumber \\
 \yDr &=& -\xDl \cos\theta_D
\sin\theta_D \left[ \frac{1}{(p_{\chi}-p_D)^2 - m^2_{D_0^1}}-
\frac{1}{(p_{\chi}-p_D)^2 - m^2_{D_0^2}}\right] \, . \label{YQ}
\end{eqnarray}
\end{widetext}

All outgoing momenta are assumed to be physical momenta. In the rest
frame of the initial state $\DQ_{1/2}$ we therefore have the
products
\begin{equation}
(p_{u^c} \cdot p_D) = M_D E_u \end{equation}
\begin{equation} (p_{d^c} \cdot p_D) = M_D E_d \end{equation}
\begin{equation} (p_{\chi} \cdot p_D) = M_D E_{\chi} = M_D^2 - M_D(E_d + E_u)
\end{equation}
\begin{equation} (p_{u^c} \cdot p_{d^c}) = \frac{1}{2}M_{\chi}^2 - \frac{1}{2}M_D^2 + M_D(E_u + E_d) \end{equation}
\begin{equation} (p_{u^c} \cdot p_{\chi}) = \frac{1}{2}M_{D}^2 - \frac{1}{2}M_{\chi}^2 - M_D E_d \end{equation}
\begin{equation} (p_{d^c} \cdot p_{\chi}) = \frac{1}{2}M_{D}^2 -
\frac{1}{2}M_{\chi}^2 - M_D E_u \, .
\label{dotproducts} \end{equation}
The decay rate can be computed by integrating the partial width
\begin{equation}
\diff \Gamma = \frac{M_D}{256 \pi^3} |\oline{\mathcal{M}}|^2 \diff
x_d \diff x_u \label{dGamma} \end{equation}
where $x_i = 2E_i/M_D$. The numerical integration can be performed
for $m_u = m_d = 0$ (but retaining $m_{\chi} \neq 0$) by following
standard techniques~\cite{Barger}.

For massless quarks, there is no interference between the diagrams
involving the $\lambda^{10}$ vertex and those from the $\lambda^9$
vertex. The latter may be obtained from the above by the
substitutions
\begin{eqnarray}
|\lambda^{10}|^2 &\to& 4|\lambda^9|^2 \nonumber \\
\cos\theta_D & \leftrightarrow& \sin\theta_D \nonumber \\
X_{iL} &\leftrightarrow & -X_{iR} \nonumber \\
m_{\tilde{q}_R}^2 &\to& m_{\tilde{q}_L}^2 \, . \end{eqnarray}
Thus, for example, we have
\begin{eqnarray} \yDr \to \zDl, &\quad& \yDl \to \zDr \nonumber \\
\ydr \to \zdl, &\quad& \yur \to \zul \, ,
\end{eqnarray}
where
\begin{equation}
\zdl = -\frac{\xdl}{(p_{u^c}-p_D)^2 - m^2_{\tilde{d}_L}}
\end{equation}
\begin{equation} \zul = -\frac{\xul}{(p_{d^c}-p_D)^2 - m^2_{\tilde{u}_L}} \end{equation}
\begin{equation} \zDl = -\xDl \left[ \frac{\cos^2\theta_D}{(p_{\chi}-p_D)^2 -
m^2_{D_0^1}}+ \frac{\sin^2\theta_D}{(p_{\chi}-p_D)^2 -
m^2_{D_0^2}}\right] \end{equation}
\begin{eqnarray} \zDr &=& +\xDr \cos\theta_D \sin\theta_D \times
\nonumber \\
 & & \left[
\frac{1}{(p_{\chi}-p_D)^2 - m^2_{D_0^1}}-
 \right. \nonumber
\\
& & \left.
\frac{1}{(p_{\chi}-p_D)^2 -  m^2_{D_0^2}}\right] \, .
\label{Zdef} \end{eqnarray}

For leptoquark processes we distinguish between rates to $u\,
e$ and $d\, \nu$ final states. In analogy to the $Y$ and $Z$ factors
above, we define the quantities
\begin{equation}
\qel = -\frac{\xel^*}{(p_{u}-p_D)^2 -
m^2_{\tilde{e}_L}}\end{equation}
\begin{equation} \qul = -\frac{\xul^*}{(p_{e}-p_D)^2 - m^2_{\tilde{u}_L}} \end{equation}
\begin{equation} \rnl = -\frac{\xnl^*}{(p_{d}-p_D)^2 - m^2_{\tilde{\nu}_L}}\end{equation}
\begin{equation} \rdl = -\frac{\xdl^*}{(p_{\nu}-p_D)^2 - m^2_{\tilde{d}_L}} \end{equation}
\begin{equation} \ser = \frac{\xer^*}{(p_{u}-p_D)^2 - m^2_{\tilde{e}_R}} \end{equation}
\begin{equation} \sur = \frac{\xur^*}{(p_{e}-p_D)^2 - m^2_{\tilde{u}_R}} \, .
\label{QRS} \end{equation}
The appropriate three-body decay rates can then be found following
the above description using the substitution rules summarized in
Table~\ref{3body}

\begin{table}[thb]
\begin{center}
\begin{tabular}{|ccccc|}
\hline  $\Gamma_{10}$ &$\Gamma_{9}$ &$\Gamma_{7}^{ue}$
&$\Gamma_{7}^{d\nu}$
&$\Gamma_{6}$ \\
$D\rightarrow u^c d^c \chi$ & $D\rightarrow u^c d^c \chi$ &$D
\rightarrow u e \chi$ & $D \rightarrow d \nu \chi$ & $D \rightarrow
u e \chi$  \\ \hline
$C |\lambda^{10}|^2$&$4 C |\lambda^{9}|^2$&$ |\lambda^{7}|^2$&$ |\lambda^{7}|^2$&$ |\lambda^{6}|^2$\\
$p_{d^c}$ & $p_{d^c}$ &$p_{e}$ &$p_{\nu}$ &$p_{e}$ \\
$p_{u^c}$ & $p_{u^c}$ &$p_{u}$ &$p_{d}$ &$p_{u}$ \\
$X_{d_R}$ &$-X_{d_L}$ &$-X_{e_L}^*$ &$-X_{\nu_L}^*$ &$X_{e_R}^*$ \\
$X_{u_R}$ &$-X_{u_L}$ &$-X_{u_L}^*$ &$-X_{d_L}^*$ &$X_{u_R}^*$ \\
$X_{D_L}$ &$-X_{D_R}$ &$X_{D_L}$ &$X_{D_L}$ &$-X_{D_R}$\\
$X_{D_R}$ &$-X_{D_L}$ &$X_{D_R}$ &$X_{D_R}$ &$-X_{D_L}$\\
$\cos\theta_D$& $\sin\theta_D$& $\cos\theta_D$& $\cos\theta_D$& $\sin\theta_D$\\
$\sin\theta_D$& $\cos\theta_D$& $\sin\theta_D$& $\sin\theta_D$& $\cos\theta_D$\\
$m^2_{\tilde{d}_R}$&$m^2_{\tilde{d}_L}$&$m^2_{\tilde{e}_L}$&$m^2_{\tilde{\nu}_L}$&$m^2_{\tilde{e}_R}$\\
$m^2_{\tilde{u}_R}$&$m^2_{\tilde{u}_L}$&$m^2_{\tilde{u}_L}$&$m^2_{\tilde{d}_L}$&$m^2_{\tilde{u}_R}$\\
$Y_{D_L}$&$Z_{D_R}$&$Y_{D_L}$&$Y_{D_L}$&$Z_{D_R}$\\
$Y_{D_R}$&$Z_{D_L}$&$Y_{D_R}$&$Y_{D_R}$&$Z_{D_L}$\\
$Y_{d_R}$& $Z_{d_L}$&$ \qel $& $\rnl$ & $ \ser$ \\
$Y_{u_R}$& $Z_{u_L}$& $\qul$ & $\rdl $&$ \sur $\\
\hline
\end{tabular}
\end{center}
{\caption{\label{3body}\footnotesize {\bf Substitution rules for
obtaining three-body decay rates}. For each process, replace the
quantity in~(\ref{msquared})~-~(\ref{dotproducts}) with the
appropriate variable from the table. }}
\end{table}

\clearpage


\end{document}